\documentclass[aps,prb,reprint,superscriptaddress]{revtex4-2}
\usepackage{amsmath,amssymb} % mathematics
\usepackage{amsmath} % mathematics
\usepackage{amsfonts} % mathematics
\usepackage{bm} % bold in math-mode
\usepackage[dvipdfmx]{graphicx} % figure
\usepackage{braket} % braket
\usepackage{physics}
\usepackage{color}
\usepackage{comment}
\usepackage{mathtools}

\begin{document}
\title{Spinor ice correlation in flat-band electronic states on kagome and pyrochlore lattices with spin-orbit coupling}

% Author information
\author{Hiroki Nakai}
\email{nakai-hiroki3510@g.ecc.u-tokyo.ac.jp}
\affiliation{Graduate School of Arts and Sciences, University of Tokyo, Meguro-ku, Tokyo 153-8902, Japan}

\author{Masafumi Udagawa}
\affiliation{Department of Physics, Gakushuin University, Mejiro, Toshima-ku, Tokyo 171-8588, Japan}

\author{Chisa Hotta}
\affiliation{Graduate School of Arts and Sciences, University of Tokyo, Meguro-ku, Tokyo 153-8902, Japan}

\date{\today}

\begin{abstract} 
We investigate the emergence and transformation of pinch-point singularities in the excitation spectrum 
of electronic flat-band systems on kagome and pyrochlore lattices with spin-orbit coupling (SOC) and Coulomb interactions. 
While pinch points are widely recognized as signatures of classical spin liquids, 
they also appear in electronic flat-band systems when there is a singular band-touching point to dispersive bands. 
We explore how SOC modifies the pinch-point structure in the chiral spin flat-band metallic state, 
which we term spinor-ice. 
The pinch point profile can rotate or redistribute its spectral weight, 
governed by a prefactor in the spectral function that primarily depends on the direction of 
the ground-state spin polarization. 
We show that the spinor-ice state could be experimentally probed by 
rotating the spin polarization of the injected electron to infer internal magnetic structures. 
These observations are discussed in conjunction with angle-resolved photoemission spectroscopy 
and the application to the potential SOC flat-band material CsW$_2$O$_6$. 
We also demonstrate the persistent residual pinch-point features 
under Coulomb interactions and deviations from the ideal flat-band limit.
\end{abstract} 

\maketitle

\section{Introduction}
%%%%%%%%%%%%%%%%%%%%%%%%%%%%%%%
%flat band physics
Flat band systems have served as an ideal platform for exploring many exotic quantum states and intriguing collective behaviors that are promoted by the quenching of kinetic energy and many-body effects \cite{Regnault2022,Balents2020,Checkelsky2024}. 
Typical examples include the occurrence of ferromagnetism \cite{Mielke1991, Tasaki1992, Mielke_Tasaki1993} and Wigner crystallization \cite{DasSarma2007}, whose spin and charge configurations are dictated by electronic interactions. 
Recent discoveries extend the relevance of flat bands beyond perfectly non-dispersive scenarios. 
A notable instance is the superconductivity observed in the twisted-bilayer graphene \cite{Cao2018_1, Cao2018_2}, 
where the two layers at a specific ``magic" angle generate nearly flat bands favorable for electron pairing \cite{Bistritzer2011}. 
Fractional Chern numbers in spin-orbit coupled (SOC) bands can appear only 
when the bands have finite dispersiveness, mimicing the quasi-quantum Hall effect in a zero-field \cite{Tang2011, Sun2011, Neupert2011,Sheng2011,Wang2011,Regnault2011}. 
%flat band references: Xie2019,Mao2020,Kang2020_FeSn,Kang2020_CoSn,Lin2018
\par
When we focus on perfect flat bands based on the mathematical construction of 
a line-graph family of lattices \cite{Mielke1991, Tasaki1992, Miyahara2005, Liu2022, Calugaru2022,Katsura_2010} like kagome and pyrochlore, 
their underlying topology deserves even larger attention~\cite{Sutherland1986,Bergman2008,Maimaiti2017,Rhim2019,Hwang2021_1,Hwang2021_2,Hwang2021_3,Rhim2020,Rhim2021,Maimaiti2017,Maimaiti2019,Mizoguchi2019,Maimaiti2021,Essafi_2017,Bilitewski2018}. 
In a one-particle eigenstate of a flat band, 
the electron occupies a few sites in real space that form a closed loop. 
The electron is quantum mechanically forbidden to hop outside the loop, 
which is described by the destructive interference condition. 
Such loop states can take two distinct forms: compact localized states (CLS), 
which are confined to small regions such as hexagons, and non-contractible loop states (NLS), 
which wind around the entire system. 
A set of CLS forms the majority of the flat-band Bloch states across almost the entire Brillouin zone (BZ). 
However, when a dispersive band touches the flat band at $\vb*{k}=\vb*{k^*}$, 
the Bloch wave function at the touching point is composed solely of NLS, 
which is not adiabatically connected to CLS: this is the source of the singularity. 
We have recently demonstrated a fundamental relationship between these two topologically distinct 
classes of loop states \cite{Udagawa2024}.
\par
When the aforementioned singular flat band exists, it manifests as a ``pinch point", 
a characteristic intensity modulation in momentum-resolved spectroscopy that develops around $\vb*{k^*}$. 
Originally, pinch points were recognized as experimental signatures of classical spin liquids, 
arising from the macroscopic degeneracy of the spin ice state~\cite{Bramwell2021} in insulating pyrochlore magnets. 
They were indeed observed in neutron scattering experiments on the canonical spin ice compound Ho$_2$Ti$_2$O$_7$~\cite{Fennell2009,Kadowaki2009} and many related systems~\cite{Moessner1998,Garanin1999,Canals2001,Rehn2016,Benton2021,Abhinav2018,Li1994,Isakov2015,Youngblood1980,Petit:2016aa,PhysRevX.8.021053,Ghosh:2019aa,Yan2023,PhysRevLett.128.107201,PhysRevResearch.5.L012025,PhysRevB.111.064417,PhysRevB.110.L020402}.   %and Dy$_2$Ti$_2$O$_7$~\cite{Bramwell2001,Fennell2009,Kadowaki2009}. 
The zero-energy excitations of the spin-ice state, described within the large-$N$ approximation \cite{Isakov2004,Garanin1999,Henley2005,doi:10.1139/p01-099,Sen2013,Conlon2010,PhysRevB.109.174421}, 
reveal that the corresponding bosons form flat bands. 
In this framework, the divergence-free constraint on spin orientations \cite{Isakov2004,Castelnovo2008} 
in spin ice translates into a destructive interference condition for bosonic wave functions, 
giving rise to CLS and NLS.
\par
Beyond the ideal pinch-point structure, a precursor of pinch points, %broken pinch points, 
often referred to as half-moon patterns~\cite{Conlon2010,PhysRevB.94.104416,Mizoguchi2017,Mizoguchi2018,Yan2018,Ghosh:2019aa,PhysRevResearch.5.L012025}, has been reported in spin-ice-related materials, 
including dipolar-octupolar pyrochlore Nd$_2$Hf$_2$O$_7$ \cite{Samartzis2022} 
and kagome ice material Nd$_2$Zr$_2$O$_7$ \cite{Xu2019}. 
The half-moon structure represents a cross section of the dispersive spectrum at higher energies~\cite{Mizoguchi2018}, 
as observed in Tb$_2$Ti$_2$O$_7$ \cite{Fennell2014}. 
Continuum field theory has shown that these dispersive levels originate from the curl-free 
counterpart of the divergence-free constraint in spin ice \cite{Yan2018}. 
In kagome ice, the pinch point is replaced by the half-moon pattern when longer-range magnetic interactions 
drive the clustering of topological excitations, modifying the ground state \cite{Mizoguchi2017}.
\par
Although many established findings on pinch points in magnetic systems can explain key features of 
the one-particle spectrum in electronic flat bands, there are some crucial differences, and several open questions remain. 
For example, $n$-fold pinch-point singularities in spectral intensity are, for some reason, tightly connected with the quantized Berry phase $n\pi$ integrated around $\vb*{k^*}$ \cite{Yan2023}. 
In fact, analyzing noninteracting or weakly interacting fermions with singular band touching
is much more tractable than dealing with the highly degenerate spin ice state. 
\par
In the present paper, we focus on the SOC electronic system realized 
in kagome and pyrochlore systems. 
Importantly, in addition to the charge degrees of freedom that mimic bosonic excitations in classical spin systems, 
electrons introduce intrinsic spin degrees of freedom, 
which imprint their correlations onto the pinch-point structure, 
transforming the conventional spin-ice picture into what we call ``spinor-ice". 
The Coulomb interactions naturally present in these systems can drive charge-order instabilities, 
potentially leading to the emergence of half-moon structures. 
Some of the authors previously demonstrated that these systems host singular flat bands 
for specific choices of spin-dependent electronic hopping induced by SOC \cite{Nakai2022}, 
and discussed their relevance to the 5$d$ material CsW$_2$O$_6$ \cite{Okamoto2020, Okamoto2024}. 
These findings suggest that CsW$_2$O$_6$ could serve as a realistic platform for 
experimentally manipulating pinch points via spin degrees of freedom. 
\par
In Sec.~\ref{sec:model}, we provide basic information about our model systems and formulation. 
The SOC effect in the noninteracting system is discussed in Sec.~\ref{sec:spinor-ice}. 
We further show how the spectrum is affected by the Coulomb interaction and present calculations in Sec.~\ref{sec:correlation}, 
and  examine the realistic situation to be observed in spin-selective angle-resolved photoemission spectroscopy 
(ARPES) experiments. 

%*%*%*%*%*%*%*%*%*%*%*%*%*%*%*%*%*%*%*%*%*%*%*%*%*%*%*%*%*%*%*%*%*%*%*%*%*%*%*%*%*%*%*%*%*%*%*%*%*%*%*%*%*%*%*%*%*
%*%*%*%*%*%*%*     Model      %*%*%*%*%*%*%*%*%*%*%*%*%*%*%*%*%*%*%*%*%*%*%*%*%*%*%*%*%*%*%*%*%*%*%*%*%*%*%*%*%*%*
%*%*%*%*%*%*%*%*%*%*%*%*%*%*%*%*%*%*%*%*%*%*%*%*%*%*%*%*%*%*%*%*%*%*%*%*%*%*%*%*%*%*%*%*%*%*%*%*%*%*%*%*%*%*%*%*%*
%*%*%*%*%*%*%*%*%*%*%*%*%*%*%*%*%*%*%*%*%*%*%*%*%*%*%*%*%*
\section{Model for flat bands} 
\label{sec:model}
%*%*%*%*%*%*%*%*%*%*%*%*%*%*%*%*%*%*%*%*%*%*%*%*%*%*%*%*%*
\begin{figure*}[tbp]
\begin{center}
	\includegraphics[width=17cm]{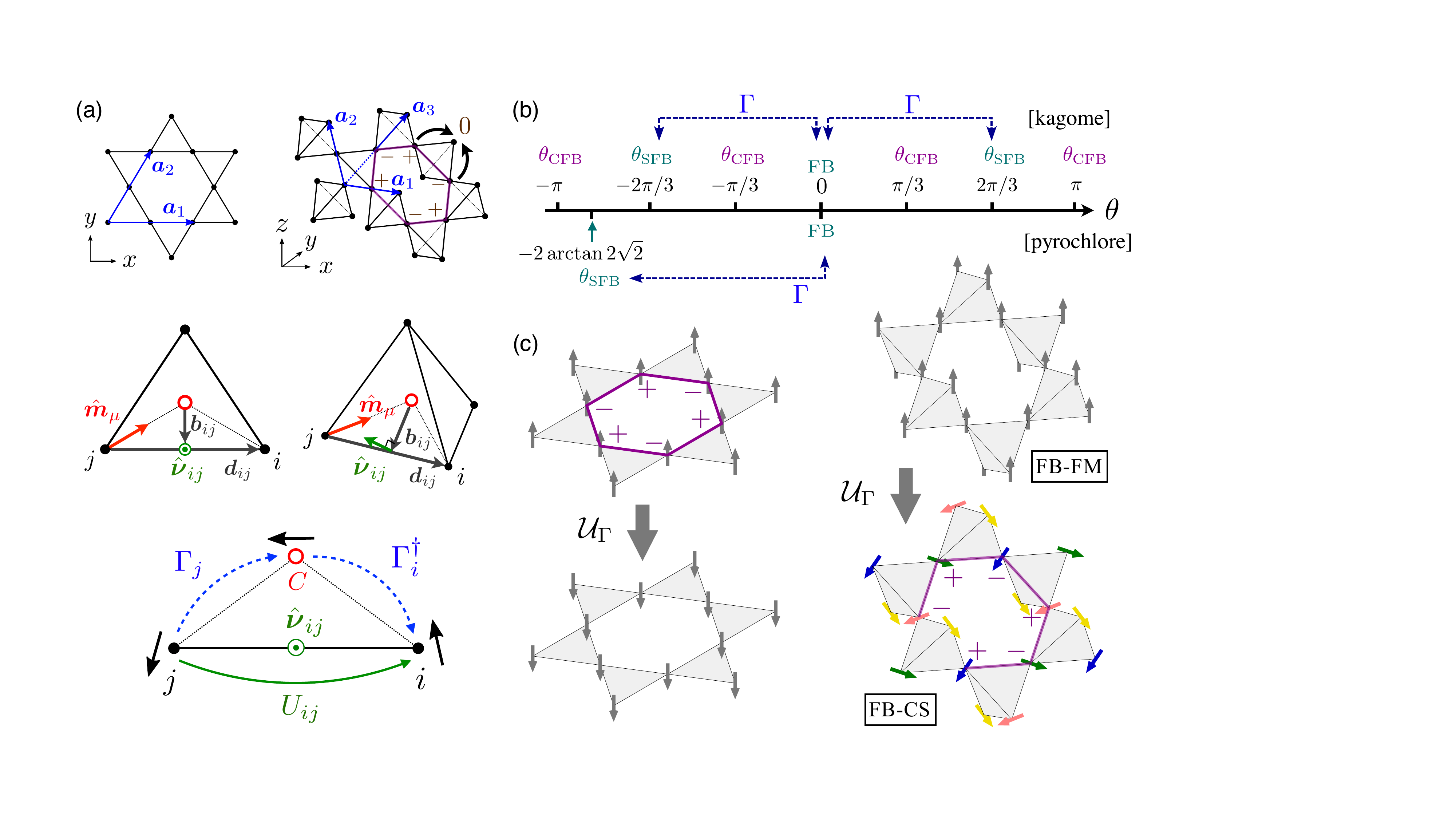}
	\caption{
	(a) Kagome and pyrochlore lattices. 
        Illustration of hopping of electrons from site $j$ to $i$, 
         accompanying the SU(2) spin rotation 
         by angle $\theta$ about the $\hat{\vb*{\nu}}_{ij}$ axis 
         shown for triangle and tetrahedron units of two lattices. 
         In the center panel, the electron hops twice via the red site at the center of 
         the unit triangle/tetrahedron by the operator $\Gamma_j$, 
         that rotates its spin orientation by angle $\pi$ about $\hat{\vb*{m}}_j$ axis. 
         When $\theta_{\rm SFB}$, it replaces the direct hopping represented by $U_{ij}$. 
         (b) Diagram of band structures as a function of $\theta$ for the kagome (upper) and pyrochlore lattice (lower side). 
         For angles $\theta_{\rm SFB}$, the flat band appears at the band bottom, 
         and for $\theta_{\rm CSB}$ the flat band appears at the center, where we have a chiral symmetry. 
         The Bloch Hamiltonians at $\theta=0$ and $\theta_{\rm SFB}$ are related by the gauge transformation using 
         the operator $\Gamma$. 
         (c) Schematic illustration of the ground states when the flat band is half filled. 
	The FB-CS state is given by the unitary transformation $\mathcal{U}_{\Gamma}$ from the FB-FM state.
}
	\label{f1}
	\end{center}
\end{figure*}
%*%*%*%*%*%*%*%*%*%*%*%*%*%*%*%*%*%*%*%*%*%*%*%*%*%*%*%*%*
%
\subsection{Model Hamiltonian}
We consider the Hamiltonian of electrons given by $\mathcal{H}=\mathcal{H}_{\rm kin}+\mathcal{H}_{\rm int}$ with 
\begin{eqnarray}
\label{eq:Ham_kin} 
\mathcal{H}_{\rm kin} &=&  \sum_{\langle i,j \rangle} \sum_{\alpha\beta} \big(-t \delta_{\alpha\beta} +i\lambda (\hat{\vb*{\nu}}_{ij} \cdot\vb*{\sigma})_{\alpha\beta} \big) c^\dagger_{i\alpha} c_{j\beta},  \\
\label{eq:Ham_int} \mathcal{H}_{\rm int} &=& U\sum_i n_{i\uparrow} n_{i\downarrow} +V\sum_{\langle i,j \rangle} n_i n_j, 
\end{eqnarray}
where $c^\dagger_{i\sigma}$ ($c_{i\sigma}$) creates (annihilates) an electron at site $i$ with spin $\sigma=\uparrow, \downarrow$, and $n_i =n_{i\uparrow}+n_{i\downarrow}$ with $n_{i\sigma} =  c^\dagger_{i\sigma} c_{i\sigma}$ represents an electron number at site $i$; $\langle i, j \rangle$ denotes the nearest-neighbor pair; 
$\vb*{\sigma} = (\sigma_x, \sigma_y, \sigma_z)$ are Pauli matrices. 
We have both the on-site $U$ and the nearest-neighbor $V$ interactions. 
\par
The kagome and pyrochlore lattices are shown in Fig.~\ref{f1}(a) whose detailed parameters are given 
in Appendix~\ref{app:pyrochlore}. 
We also consider the hyperkagome lattice obtained by depleting 1/4 of the sites regularly from the 
pyrochlore lattice. 
The kinetic part (\ref{eq:Ham_kin}) has both the spin-independent and the SOC-induced spin-dependent hopping term, whose hopping amplitudes are denoted by $t$ and $\lambda$, respectively.  
The vector $\hat{\vb*{\nu}}_{ij}$ included in the spin-dependent one is expressed as $\hat{\vb*{\nu}}_{ij} = \vb*{b}_{ij} \cross \vb*{d}_{ij} /|\vb*{b}_{ij} \cross \vb*{d}_{ij}|$ with $\hat{\vb*{\nu}}_{ij} = -\hat{\vb*{\nu}}_{ji}$, where $\vb*{b}_{ij}$ points from the center of tetrahedron/triangle (pyrochlore/kagome) to the bond center, and $\vb*{d}_{ij}$ points along the bond as $j\rightarrow i$, as shown in Fig.~\ref{f1}(a). 
This expression is determined by the lattice symmetry of the pyrochlore lattice \cite{Kurita2011}. 
On the other hand, for the kagome lattice, $\hat{\vb*{\nu}}_{ij}$ allowed by the lattice symmetry is not uniquely determined, but the possible expressions are related by the local gauge transformation~\cite{Kim2015}. 
\par
The role of the spin-dependent hopping term can be illustrated by rewriting Eq.~(\ref{eq:Ham_kin}) as 
\begin{equation}
\label{eq:Ham_kin2}
\mathcal{H}_{\rm kin} = -t_{\rm eff} \sum_{\langle i,j \rangle} \vb*{c}^\dagger_i U_{ij}(\theta) \vb*{c}_j, 
\end{equation}
with $\vb*{c}_i = ( c_{i\uparrow}, c_{i\downarrow} )^{\rm T}$ and $t_{\rm eff}=\sqrt{t^2 +\lambda^2}$. 
An SU(2) matrix $U_{ij}(\theta) = e^{ -i\theta\hat{\vb*{\nu}}_{ij}\cdot\vb*{\sigma} /2}$ rotates the electron spin by angle $\theta=2\arctan(\lambda/t)$ about the $\hat{\vb*{\nu}}_{ij}$-axis when it hops from site $j$ to $i$, and this angle varies with $\lambda/t$. 
We set $t_{\rm eff}=\sqrt{t^2 +\lambda^2}=1$ as an energy unit. 
%
%*%*%*%*%*%*%*%*%*%*%*%*%*%*%*%*%*%*%*%*%*%*%*%*%*%*%*%*%*%*%*%*
\subsection{Lattice and electronic states}
Let the lattice vector be given as $\vb*{R}_l=\sum_{\mu} l_\mu \vb*{a}_\mu,\; l_\mu \in {\mathbb{Z}}$ with $\vb*{a}_\mu$ being the unit vector of the lattice coordinate. 
The reciprocal lattice vectors are 
$\vb*{G}= \sum_{\mu} g_\mu \vb*{b}_\mu$ with integer $g_\mu$ and 
$\vb*{a}_\mu\cdot \vb*{b}_\nu=2\pi \delta_{\mu\nu}$. 
In the kagome and pyrochlore lattices, we set the sublattice vectors as 
\begin{equation}\label{vec:sublat}
\vb*{r}_\mu =
\begin{cases}
0 & (\mu=0), \\
\frac{1}{2}\vb*{a}_{\mu} & (\mu\neq 0),
\end{cases}
\end{equation}
with $\mu=0,\cdots,n_s-1$ where $N=n_sN_c$. 
By reading off the site index $j=1,\cdots,N$ as $(l,\mu)$, we apply a Fourier transformation as 
\begin{equation}\label{eq:fourier}
c_{\vb*{k}\mu\sigma} =\frac{1}{\sqrt{N_c}} \sum_{l=0}^{N_c-1} e^{-i\vb*{k}\cdot \vb*{R}_l} c_{l \mu \sigma}, 
\end{equation}
for $\vb*{c}_{\vb*{k}}=(c_{\vb*{k}0\uparrow},c_{\vb*{k}0\downarrow}, \ldots, 
c_{\vb*{k}n_s-1\uparrow}, c_{\vb*{k}n_s-1\downarrow} )^{\rm T}$, and the Hamiltonian is rewritten as
\begin{equation}\label{eq:Ham_kin3}
\mathcal{H}_{\rm kin} = \sum_{\vb*{k}} \vb*{c}^\dagger_{\vb*{k}} \mathcal{H}(\vb*{k},\theta) \vb*{c}_{\vb*{k}},
\end{equation}
which satisfy 
\begin{equation}\label{eq:ham_theta}
\begin{split}
& \mathcal{H}(\vb*{k},\theta) \vb*{u}_{n} (\vb*{k}) = \varepsilon_n(\vb*{k}) \vb*{u}_{n} (\vb*{k}). 
\end{split}
\end{equation}
The $n$-th Bloch state is expanded as 
\begin{equation}\label{eq:blochstate}
c^\dagger_{n\vb*{k}}= \sum_{\mu,\sigma} [\vb*{u}_n(\vb*{k})]_{\mu\sigma} c^\dagger_{\vb*{k}\mu\sigma}. 
\end{equation}
In the presence of time-reversal symmetry and inversion symmetry, all eigenenergies are at least twofold degenerate. 
\par
%*%*%*%*%*%*%*%*%*%*%*%*%*%*%*%*%*%*%*%*%*%*%*%*%*%*%*%*%*
\par
We start from the case when SOC is absent ($\theta=0$). 
The Bloch Hamiltonian is diagonal in spin space, which follows 
\begin{eqnarray}
\label{eq:ham_theta0} &&\mathcal{H}(\vb*{k}, \theta=0)=h(\vb*{k})\otimes I_{2\times 2}, \\
\label{eq:fbblchwf} && h(\vb*{k}) \vb*{v}_{n} (\vb*{k}) = \varepsilon_n(\vb*{k}) \vb*{v}_{n} (\vb*{k}), 
\end{eqnarray}
where $\vb*{v}_{n} (\vb*{k})$ of dimension $n_s$ is related to Eq.~(\ref{eq:ham_theta}) for up and down spin states as
\begin{equation}
\vb*{u}^{(0)}_{n\sigma} (\vb*{k})=\vb*{v}_n (\vb*{k})\otimes\ket*{\sigma}
\end{equation}
%*%*%*%*%*%*%*%*%*%*%*%*%*%*%*%*%*%*%*%*%*%*%*%*%*%*%*%*%*
\subsection{Flat bands}
For the kagome and pyrochlore lattices with $n_s=3$ and 4, respectively, they have $n_{\rm FB}=2$ and 4-fold degenerate flat bands at the highest level. 
In the most heuristic explanation, the flat band for $\theta=0$ is realized under the so-called ``divergence-free" condition:  
If the weight of one-body wave function has the distribution of $+1$, $-1$, and 0 for all triangular units, it is the eigenstate with zero energy dispersion and energy $+2t$. 
The most familiar one is the CLS, e.g., the one defined around a hexagonal loop with alternating $+1$ and $-1$ weights. 
Even if the electron wants to hop out of these loops, the contributions from the neighboring sites along the loop cancel out, and the electron is confined, and accordingly the dispersion is lost. 
For the SOC flat band, the same context holds, but this time, the relative angles of spin orientations of $n_s$ sublattice are quenched. 
Along the CLS, the electrons hop outside the loop by rotating their spins and cancel out with each other when the spin angles have particular relationships. 
A more rigorous explanation is given in Ref.~[\onlinecite{Nakai2022}]. 
\par
How the other flat bands appear when we introduce SOC is summarized in Fig.~\ref{f1}(b). 
These band structures are systematically displayed in Ref.~[\onlinecite{Nakai2023}] and in the latter part of this paper. 
When $\theta \ne 0$, $\mathcal{H}(\vb*{k}, \theta)$ start to have block off-diagonal elements about up and down spins, but all the energy bands are at least twofold degenerate because of the inversion and time-reversal symmetry. 
For the kagome lattice, another flat band is found at the middle level, $\varepsilon_{\rm CFB}=0$, when $\theta_{\rm CFB}= \pm \pi/3, \pm \pi$. 
At this parameter, the energy band structure has a chiral symmetry (up-side-down) \cite{Nakai2023}, which is the reason why we call it a chiral flat band (CFB). 
Pyrochlore lattice also exhibits a chiral symmetry for $\theta=-2\arctan(1/\sqrt{2}), 2\arctan\sqrt{2}$, but only part of the energy bands become flat along the symmetric line. 
When we increase $\theta$ further, the flat band is realized at the lowest level; $\theta_{\rm SFB}= \pm 2\pi/3$ and $\theta_{\rm SFB}= -2\arctan2\sqrt{2}$ for the kagome and pyrochlore lattices, respectively. 
We call them spin-orbit coupled flat band (SFB). 
%*%*%*%*%*%*%*%*%*%*%*
\subsection{Gauge transformation}
A remarkable aspect of the present model is to allow a gauge transformation $\Gamma$ that exactly relates the ones realized for $\theta=0$ and $\theta_{\rm SFB}$ \cite{Nakai2023}. 
Let us set $\theta=\theta_{\rm SFB}$, where the SU(2) gauge in Eq.~(\ref{eq:Ham_kin2}) is rewritten as 
\begin{eqnarray}
&& U_{ij}(\theta_{\rm SFB})= -\Gamma^\dagger_i \Gamma_j , 
\label{eq:uijg}\\ 
&& \Gamma_j = -i\hat{\vb*{m}}_\mu \cdot \vb*{\sigma} = \exp\left[-i\frac{\pi}{2} \hat{\vb*{m}}_\mu \cdot \vb*{\sigma}\right]. 
\label{eq:gamma}
\end{eqnarray}
As shown in Fig.~\ref{f1}(a), this factorization represents a fictitious hopping of electrons from site $j$ to the red circle site located at the center of the triangle or tetrahedron, accompanying the rotation of its spin angle by $\pi$ about the $\hat{\vb*{m}}_\mu$ axis, and to site $i$ again by another $\pi$ rotation.  
The red sites added here form a parent lattice, namely a honeycomb/diamond lattice for kagome/pyrochlore lattice, and the latter is the line graph of the former. 
This kind of factorization is possible only when $\theta=\theta_{\rm SFB}$. 
\par
Substituting Eq.~(\ref{eq:uijg}) to the kinetic Hamiltonian in Eq.~(\ref{eq:Ham_kin2}) we reach the following transformation: 
\begin{equation}\label{eq:Ham_kin4}
\mathcal{H}_{\rm kin} = +t \sum_{\langle i,j \rangle} \vb*{c}^\dagger_i \Gamma^\dagger_i \Gamma_j \vb*{c}_j = +t \sum_{\langle i,j \rangle} \tilde{\vb*{c}}^\dagger_i \tilde{\vb*{c}}_j,
\end{equation}
where we transformed the local basis as $\tilde{\vb*{c}}_i = (\tilde{c}_{i\uparrow}, \tilde{c}_{i\downarrow})^{\rm T}= \Gamma_i \vb*{c}_i$, and the effect of SOC of rotating the spins is renormalized into the local gauge. 
\par
The above transformation can be rewritten as, $\Gamma_i U_{ij} \Gamma^\dagger_j = -I$, and by introducing the global operator as their product, $\Gamma=\oplus_{\mu=1}^{n_s} \Gamma_\mu$, the two flat band Hamiltonian is related to each other as
\begin{equation}
\Gamma \mathcal{H}(\vb*{k}, \theta_{\rm SFB}) \Gamma^\dagger = -\mathcal{H}(\vb*{k}, \theta=0). 
\end{equation}
Accordingly, the eigenstates of $\mathcal{H}(\vb*{k}, \theta_{\rm SFB})$ are given by transforming Eq.~(\ref{eq:fbblchwf}) as 
\begin{equation}\label{eq:fbblchwf2}
\tilde{\vb*{u}}_{n\tilde{\sigma}}(\vb*{k})=\Gamma \vb*{u}^{(0)}_{n\sigma} (\vb*{k}) 
\end{equation}
with their energy changing its sign, $\varepsilon_{\rm SFB}=-\varepsilon_{\rm FB}$. 
We set their pseudospin orientations as indexed by $\tilde{\sigma}=\Uparrow,\Downarrow$ corresponding to $\sigma=\uparrow,\downarrow$. 
These pseudospins are not the actual orientation of each spin but are generated by $\Gamma_i$ in the sublattice-dependent directions, which are depicted in Fig.~\ref{f1}(c). 
Because of Kramers degeneracy, there is a degree of freedom in the choice of eigenstates for each $\vb*{k}$. 
Namely, any linear combination of $\tilde{\vb*{u}}_{n\Uparrow}(\vb*{k})$ and $\tilde{\vb*{u}}_{n\Downarrow}(\vb*{k})$ yields a set of twofold degenerate eigenstates, while this operation should be done simultaneously over the whole energy bands. 
%
%*%*%*%*%*%*%*%*%*%*%*%*%*%*%*%*%*%*%*%*%*%*
%*%*%*%*%*%*%*%*%*%*%*%*%*%*%*%*%*%*%*%*%*%*
%*%*%*%*%*%*%*%*%*%*%*%*%*%*%*%*%*%*%*%*%*%*%*%*%*%*%*%*
\subsection{Single-particle excitation}\label{sec:excitation-formula}
Measurements of the single-particle spectrum provide information on excited states within the spatial resolution of sublattice coordinates. 
We consider an electron operator forming a plane wave, expanded as  
\begin{equation}
a^\dagger_{\vb*{k}\sigma} = \frac{1}{\sqrt{N}} \sum_{l=0}^{N_c-1}\sum_{\mu=0}^{n_s-1} 
e^{i\vb*{k}\cdot(\vb*{R}_l +\vb*{r}_\mu)} c^\dagger_{l\mu\sigma}, 
\end{equation}
and evaluate the matrix element of the transition probability of adding it to or deleting it from the $N_e$-particle ground state $\ket*{\Psi^{N_e}_g}$; $P_\sigma(\vb*{k}, \omega) = P^+_\sigma (\vb*{k}, \omega) +P^-_\sigma (\vb*{k}, \omega)$ and 
\begin{equation}
P^\pm_\sigma (\vb*{k}, \omega) = \sum_{i} | \mel*{\Psi^{N_e\pm 1}_i} {a^\pm_{\vb*{k}\sigma}}{\Psi^{N_e}_g} |^2 
\delta(\hbar\omega - E^{N_e\pm 1}_i +E^{N_e}_g). 
\end{equation}
For the interacting state ${\cal H}_{\rm int}\ne 0$, we employ a Hartree-Fock approximation (see Appendix~\ref{app:pyrochlore} for details). 
We can thus write down many-body eigenstates as Slater wavefunctions composed of Bloch electrons in Eq.~(\ref{eq:blochstate}). 
The excited states that contribute to the spectral weight are in the form of $\ket*{\Psi^{N_e\pm1}_{\vb*{k}n}}=c^{\pm}_{n\vb*{k}} \ket*{\Psi^{N_e}_g}$, where the subscript $+/-$ on the operators represents the creation/annihilation operator. 
Then we find 
\begin{align}\label{eq:spectral_function}
P^\pm_\sigma (\vb*{k}, \omega)&= \sum_{\vb*{k'},n} | \mel*{\Psi^{N_e\pm 1}_{\vb*{k'}n}}{a^\pm_{\vb*{k}\sigma}}{\Psi^{N_e}_g} |^2 
\delta(\hbar\omega - E^{N_e\pm 1}_{\vb*{k'},n} +E^{N_e}_g) \nonumber\\
&= \frac{1}{n_s} \sum_{n} \Big| \sum_{\mu=0}^{n_s-1} e^{-i\vb*{k}\cdot\vb*{r}_\mu} [\vb*{u}_n(\vb*{k})]_{\mu\sigma} \Big|^2 \nonumber\\
& \quad\quad \times \delta(\hbar\omega-E^{N_e\pm 1}_{\vb*{k},n} +E^{N_e}_g). 
\end{align}
\par
We focus on the case where the Fermi level lies at the flat-band energy level. 
Then, ${\cal S}(\vb*{k})=\sum_\sigma P_\sigma(\vb*{k}, \omega=0)$ gives a structure factor of the flat-band state; 
\begin{equation}\label{eq:sk}
{\cal S}(\vb*{k}) = \frac{1}{n_s} \sum_\sigma \sum_{n\in {\rm FB}}  \bigg| \sum_{\mu=0}^{n_s-1} e^{-i\vb*{k}\cdot\vb*{r}_\mu} [\vb*{u}_n(\vb*{k})]_{\mu\sigma} \bigg|^2. 
\end{equation}

%*%*%*%*%*%*%*%*%*%*%*%*%*%*%*%*%*%*%*%*%*%*%*%*%*%*%*%*%*
%*%*%*%*%*%*%*%*%*%*%*%*%*%*%*%*%*%*%*%*%*%*%*%*%*%*%*%*%*
\begin{figure*}[t]
	\begin{center}
		\includegraphics[width=18cm]{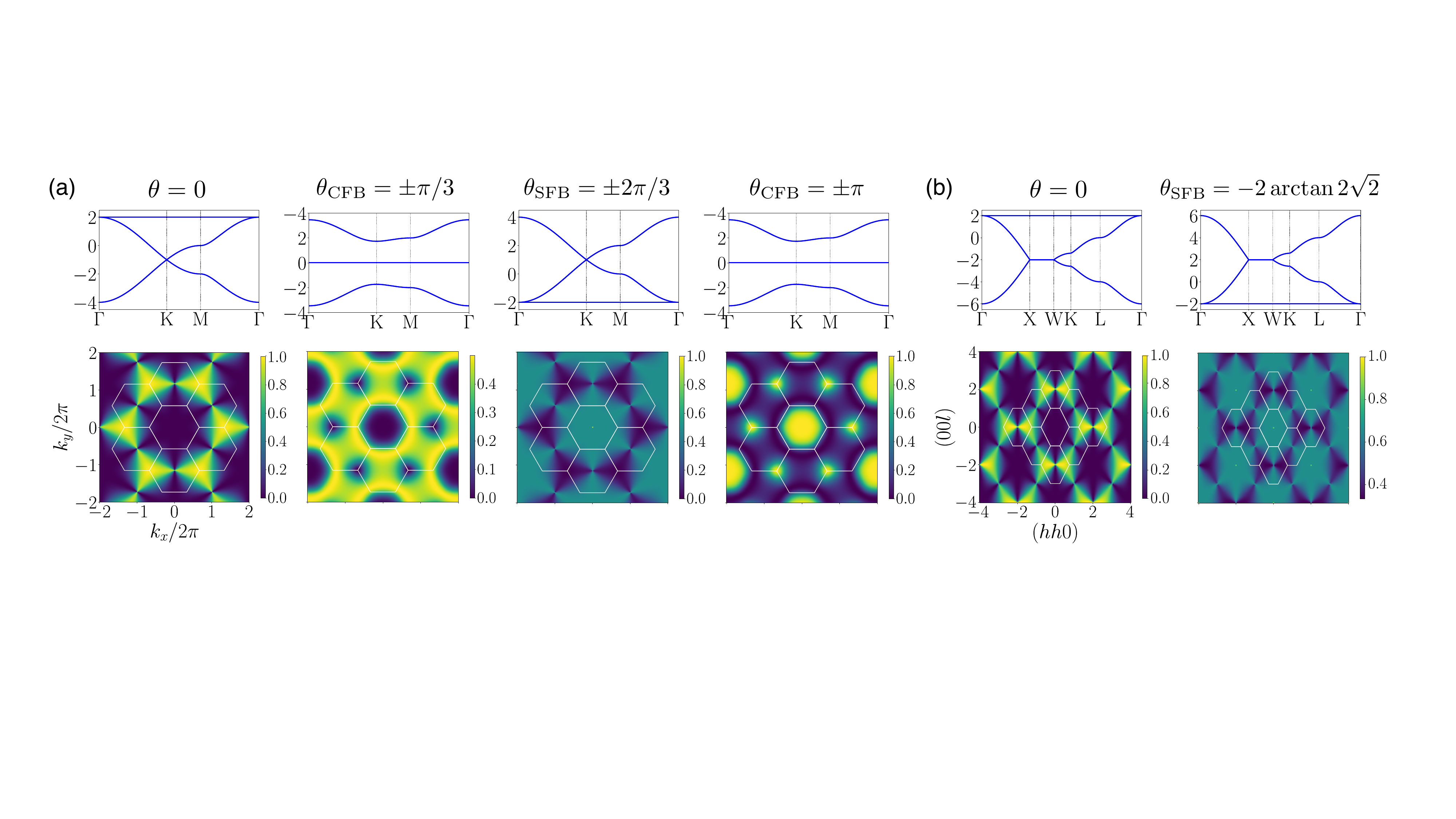}
		\caption{Nonintercting (${\mathcal H}_{\rm int}=0$) energy band structures and structure factor ${\cal S}(\vb*{k})$  
			of the noninteracting model for $U\sim 0, V=0$ obtained using Eq.~(\ref{eq:sk}). 
			(a) Kagome lattice model for four choices of $\theta$, which all host flat bands; 
			$\theta=0, \pm2\pi/3$ are the singular flat bands with NLS, and $\theta=\pm\pi/3, \pm\pi$ are nonsingular flat bands. 
			We consider the half-filled flat band as a ground state, with filling factors 
			5/6, 1/2, 1/6, and 1/2 from left to right. 
			(b) Pyrochlore lattice model for $\theta=0$ and $-2\arctan 2\sqrt{2}$. We take 3/4 and 1/4 filling, respectively. 
		}
		\label{f2}
	\end{center}
\end{figure*}
%*%*%*%*%*%*%*%*%*%*%*%*%*%*%*%*%*%*%*%*%*%*%*%*%*%*%*%*%*%*%*%*%*%*%*%*%*%*%*%*%*%*%*%*%*%*%*%*%*%*%*%*%*%*%*%*%*%*
\begin{figure}[t]
	\begin{center}
		\includegraphics[width=8.6cm]{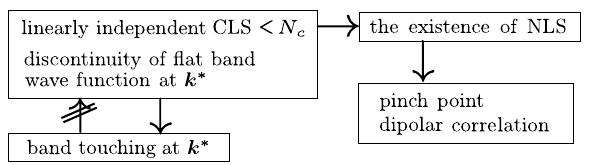}
		\caption{Illustration of the relationships between phenomena 
			related to flat band singularity. 
		}
		\label{f3}
	\end{center}
\end{figure}
%*%*%*%*%*%*%*%*%*%*%*%*%*%*%*%*%*%*%*%*%*%*%*%*%*%*%*%*%*
%*%*%*%*%*%*%*%*%*%*%*%*%*%*%*%*%*%*%*%*%*%*%*%*%*%*%*%*%*
\subsection{Noninteracting flat band states with SOC}
\label{sec:soc}
We first examine the spectral weight of the flat bands we listed in Fig.~\ref{f1}(b). 
Figure~\ref{f2}(a) shows ${\cal S}(\vb*{k})$ in Eq.~(\ref{eq:sk}) obtained for 
the flat-band states realized for four particular values of $\theta$ on a kagome lattice. 
The ones for $\theta=0, \pm 2\pi/3$ are singular flat bands that exhibit pinch-point singularity: 
a bowtie shaped gradation of intensities of ${\cal S}(\vb*{k})$ around the reciprocal points, 
$\vb*{G}=\sum_\mu g_\mu \vb*{b}_\mu$, for odd numbers of $g_1+g_2$, e.g., $g_1+g_2=1$ is the second BZ. 
When $\theta=\pm \pi /3, \pm\pi$, the pinch point disappears. 
For the pyrochlore lattice in Fig.~\ref{f2}(b), both $\theta=0$ and $-2\arctan(2\sqrt{2})$ cases are singular and have pinch points. 
\par
How the observations are related to whether the flat band is singular or not, or equivalently, whether we have pinch points or not, 
is partially clarified as shown in Fig.~\ref{f3}. 
Originally, Bergmann {\it et al.} pointed out \cite{Bergman2008} 
that the flat-band Bloch state for $\vb*{k}$ is constructed by the translational copies 
of CLS wave function $|u^{\rm CLS}_{\vb*{R}_l}\rangle$ positioned at $\vb*{R}_l$. 
For the singular flat bands, the number of independent Bloch states based on CLS is typically $N_c-1$, 
because their $N_c$-sum vanishes as $\sum_{l=0}^{N_c-1} |u^{\rm CLS}_{\vb*{R}_l}\rangle=0$. 
The missing state is the one at the band touching point $\vb*{k}$, and takes the form of Bloch states based on the translational copies of NLS 
around the periodic boundary. 
Whilst, it was shown by Rhim {\it et al.} that there is a flat band with band touching but without NLS \cite{Rhim2019}, 
namely, the band touching is a necessary but not a sufficient condition for singularity. 
For the two panels in Fig.\ref{f2}(a), $\theta=\pm\pi/3, \pm\pi$, the lack of band touching indeed destroys the pinch point; otherwise, the kagome and pyrochlore flat bands are singular. 

%*%*%*%*%*%*%*%*%*%*%*%*%*%*%*%*%*%*%*%*%*%*%*%*%*%*%*%*%*%*%*%*%*%*%*%*%*%*%*%*%*%*%*%*%*%*%*%*%*%*%*%*%*%*%*%*%*%*
%%  Interactions 
%*%*%*%*%*%*%*%*%*%*%*%*%*%*%*%*%*%*%*%*%*%*%*%*%*%*%*%*%*%*%*%*%*%*%*%*%*%*%*%*%*%*%*%*%*%*%*%*%*%*%*%*%*%*%*%*%*%*
\section{Spinor ice state}
\label{sec:spinor-ice}
In this section, we take into account the on-site Coulomb interaction for the SOC flat-band systems.
When the top or bottom flat bands are half-filled, 
the electrons redistribute within the flat-band states to minimize the Coulomb energy 
and the exact ground state is obtained by avoiding double occupancy, $\ev*{n_{i\uparrow}n_{i\downarrow}}=0$. 
The direct influence of spin degrees of freedom to the pinch point is analyzed 
based on the ``spinor ice" state picture, as we see in the following. 
%*%*%*%*%*%*%%*%*%*%*%*%*%*%*%*%*%*%*%*%*%*%*%*%*%*%*%*%*%*%*
\subsection{Ground state of SOC flat band systems}
Let us first introduce the ground state of the half-filled flat band systems with and without SOC. 
Even an infinitesimal on-site Coulomb interaction breaks the SU(2) symmetry spontaneously, 
and the ground state is ``polarized" in the $\vb*{p}$ direction; 
\begin{eqnarray}\label{eq:gs_wf}
&& \ket*{\Psi(\theta)} =  \prod_{\substack{\vb*{k}\in{\rm BZ}\\ n\in{\rm FB}}}
\Big( \sum_{\mu\sigma} [\vb*{u}_{n\vb*{p}}(\vb*{k})]_{\mu\sigma} c^\dagger_{\vb*{k}\mu\sigma}\Big) \ket{0}, \\
&& [\vb*{u}_{n\vb*{p}}(\vb*{k})]_{\mu\sigma} = \begin{dcases}
\braket*{\sigma}{\vb*{p}} [\vb*{v}_n(\vb*{k})]_\mu & (\theta=0), \\
\mel*{\sigma}{\Gamma_\mu}{\vb*{p}} [\vb*{v}_n(\vb*{k})]_\mu & (\theta=\theta_{\rm SFB}), 
\end{dcases}
\end{eqnarray}
where $\ket*{\vb*{p}}=(\cos(\theta_{\rm p}/2), \sin(\theta_{\rm p}/2)e^{i\phi_{\rm p}})^{\rm T}$. 
For $\theta=0$, it corresponds to the fully spin-polarized state called flat-band ferromagnetism (FB-FM) \cite{Mielke1991, Tasaki1992, Mielke_Tasaki1993}. 
The two ground states for $\theta=0,\theta_{\rm SFB}$ are connected by a unitary operator representing the SU(2) gauge transformation $\Gamma$; 
\begin{eqnarray}
&& \ket*{\Psi(\theta=\theta_{\rm SFB})} = \mathcal{U}_{\Gamma} \ket*{\Psi(\theta=0)}, \\
&& \mathcal{U}_{\Gamma} = \bigotimes_{l\mu} \exp\left[ i\pi \hat{\vb*{m}}_\mu \cdot\vb*{S}_{l\mu}\right], 
\end{eqnarray}
with the spin operator $\vb*{S}_{l\mu}=\sum_{\alpha\beta} c^\dagger_{l\mu\alpha}(\vb*{\sigma}_{\alpha\beta}/2)c_{l\mu\beta}$. 
Because of this SU(2) gauge transformation, the relative angles of spins among sublattices in $\ket*{\Psi(\theta_{\rm SFB})}$ are quenched as shown schematically in Fig.~\ref{f1}(c) and this state can be regarded as ``ferromagnetic" about the sublattice-dependent local quantization axis. 
This feature means that in general, $\ket*{\Psi(\theta_{\rm SFB})}$ is a flat-band chiral spin (FB-CS) state 
with finite scalar spin chirality. 
While for the kagome lattice shown in Fig.~\ref{f1}(c), 
if $\vb*{p}=\hat{\vb*{z}} \perp \hat{\vb*{m}}_\mu$, 
$\mathcal{U}_{\Gamma}$ flips all the spins upside down to another ferromagnet 
without scalar chirality.
\par
The gauge transformation $\mathcal{U}_{\Gamma}$ is safely performed because $\mathcal{H}_{\rm int}$ remains unchanged, namely, 
\begin{equation}\label{eq:Ham_int_2}
\mathcal{H}_{\rm int} = \frac{U}{2} \sum_i \vb*{c}^\dagger_i \vb*{c}_i (\vb*{c}^\dagger_i \vb*{c}_i -1) = \frac{U}{2} \sum_i \tilde{\vb*{c}}^\dagger_i \tilde{\vb*{c}}_i (\tilde{\vb*{c}}^\dagger_i \tilde{\vb*{c}}_i -1), 
\end{equation}
and we find 
\begin{equation}
\mel*{\Psi(\theta_{\rm SFB})}{n_{i\uparrow} n_{i\downarrow}}{\Psi(\theta_{\rm SFB})}=0. 
\end{equation}
\par
In evaluating Eq.~(\ref{eq:spectral_function}), the exact ground state for half-filled flat bands is available, whereas the exact one-particle/hole excited states are not feasible because the reconstruction of energy bands may occur owing to quantum many-body effect. 
However, when $U$ is small enough, adding one particle only modifies the interaction energy at most by order-1, and it is natural to expect that order-$N$ electrons remain in the flat-band state because the modification of bands will require more energy cost. 
For this reason, we employ the Hartree-Fock approximation, whose details are provided in Appendix~\ref{app:pyrochlore}. 
The $N_e+1$ excited states are obtained from the flat-band ground state by adding one electron to the $n$th empty band labeled as $\ket*{\Psi^{N_e+1}_{\vb*{k}n}}$. 
%*%*%*%*%*%*%*%*%*%*%*%*%*%*%*%*%*%*%*%*%*%*%*%*%*%*%*%*%*
\begin{figure*}[t]
	\begin{center}
	\includegraphics[width=18cm]{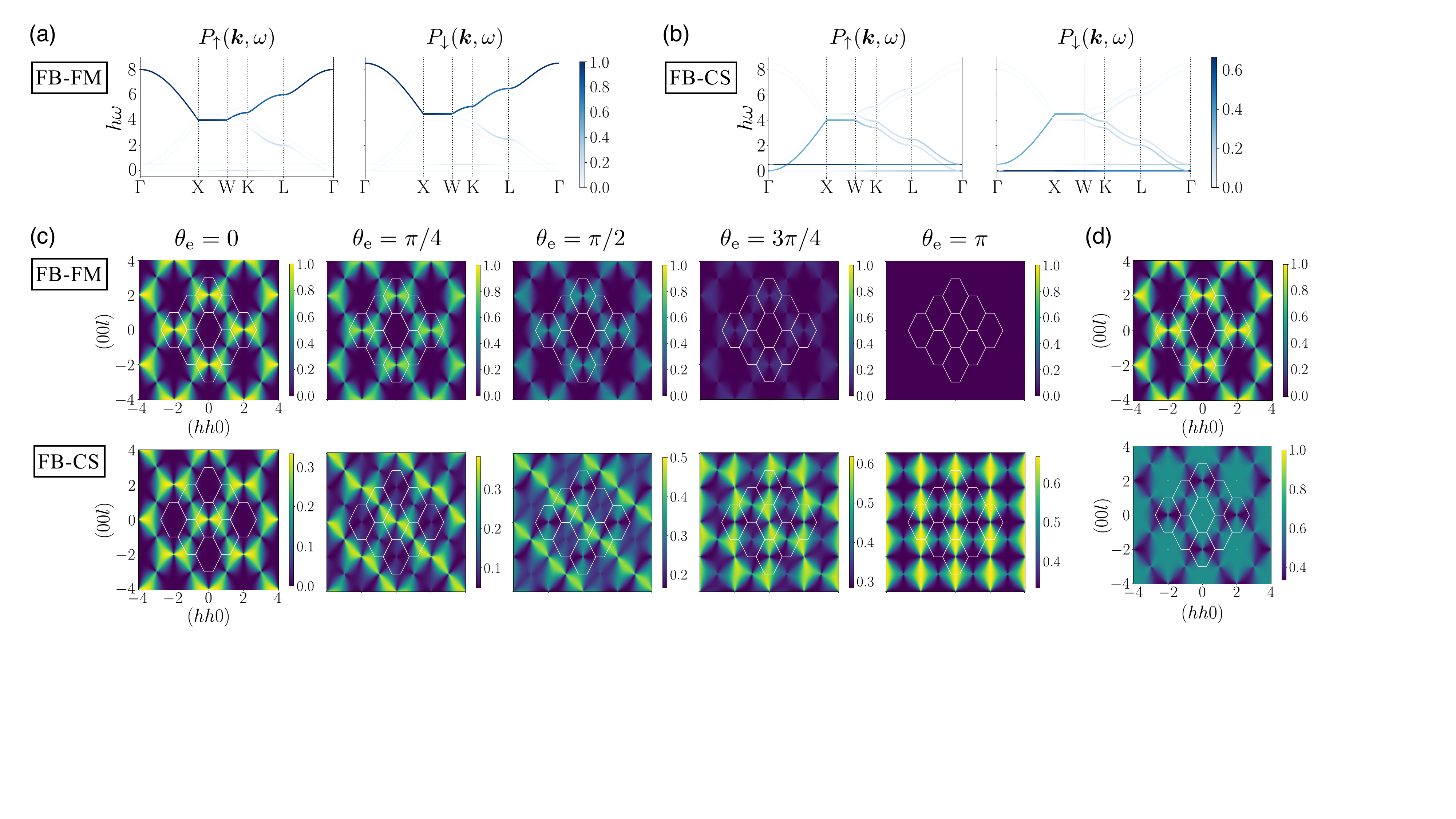}
\caption{(a)(b) Transition probability function $P_\sigma (\vb*{k}, \omega)$ 
for (a) FB-FM and (b) FB-CS states with $U/t_{\rm eff}=1$ and $V=0$ 
as the intensity plot on the plane of $\omega$ and along the path through the highly symmetric point of the first BZ. 
(c)(d) Structure factor in the $(hhl)$ plane for the FB-FM (top) and FB-CS (bottom) states; (c) spin-polarized, ${\cal S}_{\vb*{e}}(\vb*{k})$, and (d) spin-unpolarized, ${\cal S}(\vb*{k}) =\sum_{\pm\vb*{e}} {\cal S}_{\vb*{e}}(\vb*{k})$. We set the polarization of the ground state as $(\theta_{\rm p}, \phi_{\rm p})=(0,0)$ and for the spin-polarized one the polarization of the added particle as $\theta_{\rm e}=0,\pi/4,\pi/2,3\pi/4,\pi$ with $\phi_{\rm e}=0$. }
		\label{f4}
	\end{center}
\end{figure*}
%*%*%*%*%*%*%*%*%*%*%*%*%*%*%*%*%*%*%*%*%*%*%*%*%*%*%*%*%*
%*%*%*%*%*%*%%*%*%*%*%*%*%*%*%*%*%*%*%*%*%*%*%*%*%*%*%*%*%*%*%*%*%*%*%*%*%*%*%*%*%*%*%*%*%*%*%*%*%*%*%*%*%*%*%*%*%*%*%*
\begin{figure*}[t]
	\begin{center}
		\includegraphics[width=18cm]{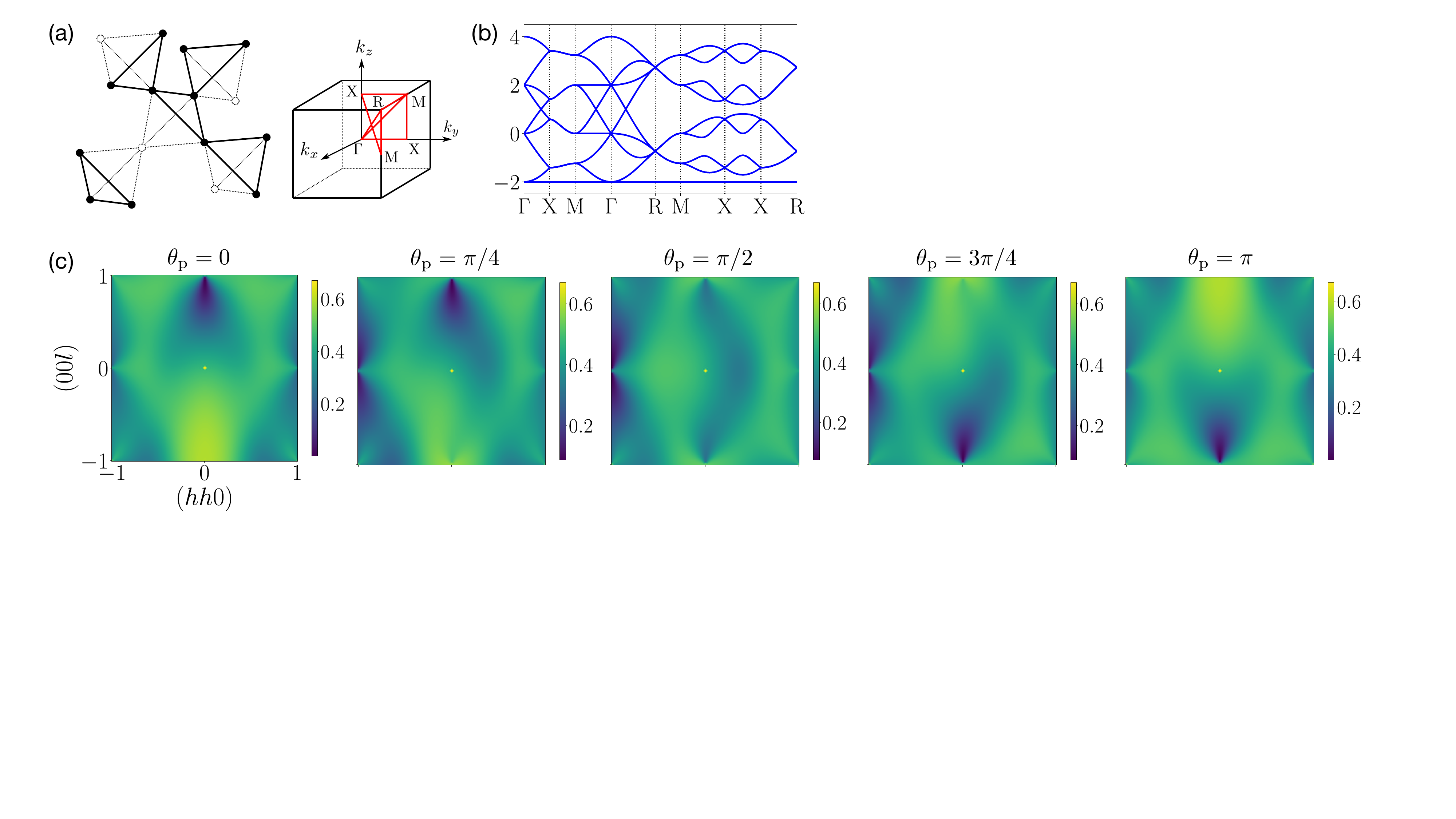}
		\caption{Hyperkagome lattice. (a) Lattice structure is represented as a 1/4-depleted pyrochlore lattice and the corresponding first BZ. (b) Band structures for $\theta_{\rm SFB}=-2\arctan2\sqrt{2}$ with the SFB for $\varepsilon=-2$. (c) Structure factor ${\cal S}(\vb*{k})=\sum_{\pm\vb*{e}} {\cal S}_{\vb*{e}}(\vb*{k})$ in the $(hhl)$ plane for the spinor ice polarized in $(\theta_{\rm p},\phi_{\rm p})$-direction. We set $\theta_{\rm p}=0,\pi/4,2/\pi,3\pi/4,\pi$ with $\phi_{\rm p}=0$. }
		\label{f5}
	\end{center}
\end{figure*}
%*%*%*%*%*%*%*%*%*%*%*%*%*%*%*%*%*%*%*%*%*%*%*%*%*%*%*%*%*
%*%*%*%*%*%*%*%*%*%*%*%*%*%*%*%*%*%*%*%*%*%*%*%*%*%*%*%*%*
\begin{figure*}[tbp]
	\begin{center}
		\includegraphics[width=17cm]{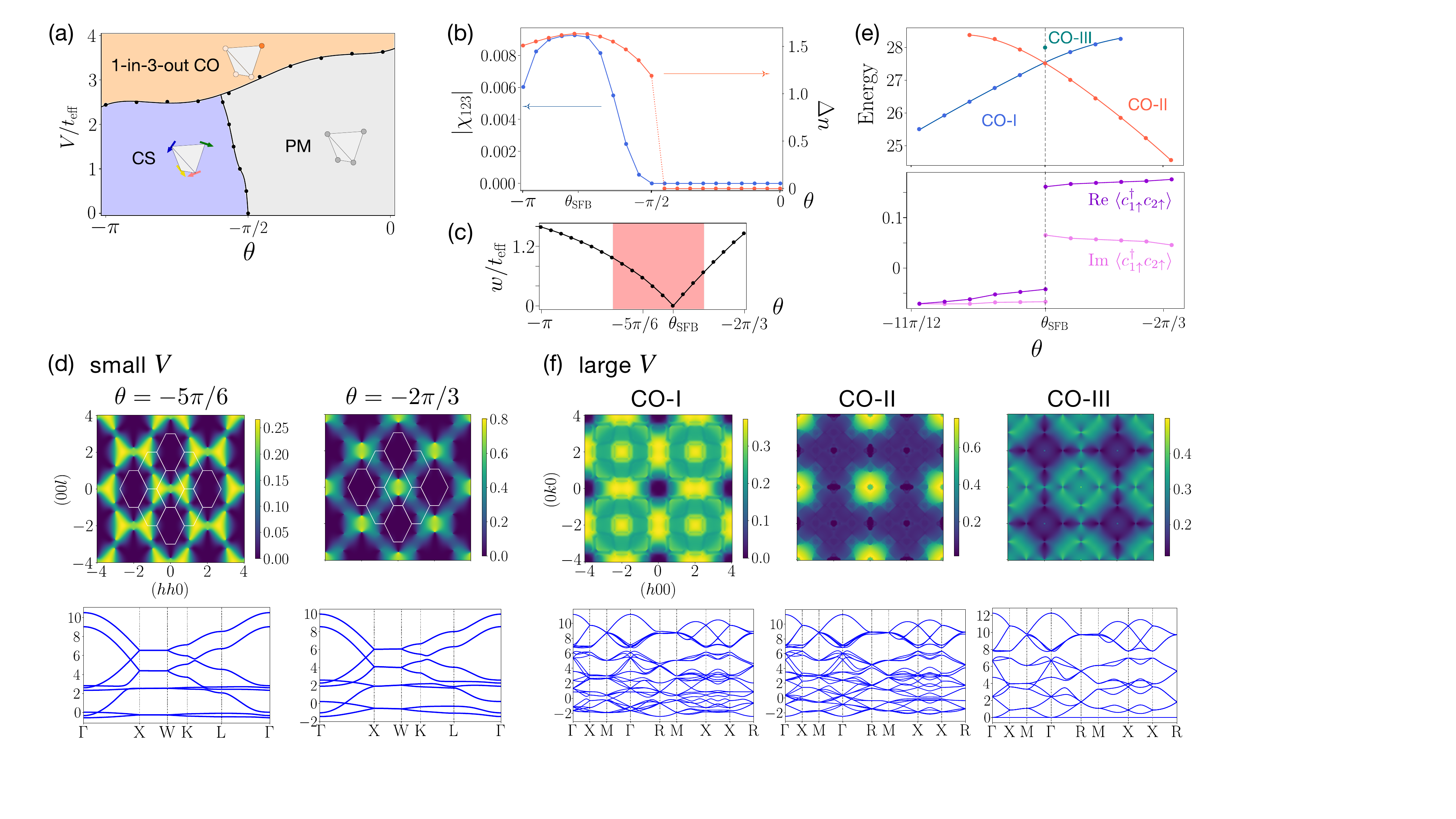}
		\caption{
(a) Phase diagram of $\mathcal{H}=\mathcal{H}_{\rm kin}+\mathcal{H}_{\rm int}$ [Eqs.~(\ref{eq:Ham_kin}) and (\ref{eq:Ham_int})]
on the $\theta$-$V/t_{\rm eff}$ plane for $U/t_{\rm eff}=5$. 
We find chiral spin (CS), paramagnetic metal (PM) and 1-in-3-out charge ordered (CO) phases. 
(b) Scalar spin chirality $\chi_{123}=\ev*{\vb*{S}_1}\cdot(\ev*{\vb*{S}_2}\cross\ev*{\vb*{S}_3})$ for $(U,V)/t_{\rm eff}=(5,1)$, 
and the charge disproportionation $\Delta n=\ev*{n_{\rm in}}-\ev*{n_{\rm out}}$ for $(U,V)/t_{\rm eff}=(5,3)$. 
(c) Bandwidth of the occupied band. The bowtie-like structure is observed in the shaded area. 
(d) Small-$V$ results: integrated spectral weight ${\cal S}(\vb*{k})$ of the occupied band in the $(hhl)$ plane and band structures for $\theta=-5\pi/6$ and $\theta=-2\pi/3$ with $(U,V)/t_{\rm eff}=(5,1)$. Here, we set the Fermi energy at zero.
(e)(f) Large-$V$ results: 3-in-1-out CO phases for $(U,V)/t_{\rm eff}=(5,4)$. 
Energies of CO-I, CO-II, and CO-III phases as a function of $\theta$, 
real and imaginary parts of the bond order parameter $\ev*{c^\dagger_{1\uparrow}c_{2\uparrow}}$ as a function of $\theta$, 
and the integrated spectral weight ${\cal S}(\vb*{k})$ of the occupied band in the $(hk0)$ plane (top) 
and band structures for $\theta_{\rm SFB}$ (bottom) for CO-I, CO-II, and CO-III phases. 
}
		\label{f6}
	\end{center}
\end{figure*}
%*%*%*%*%*%*%*%*%*%*%*%*%*%*%*%*%*%*%*%*%*%*%*%*%*%*%*%*%*
\subsection{Structure factor}
We first set $U>0$ and $V=0$ for $\theta=0$ and $\theta_{\rm SFB}$, and consider the ground state with the half-filled flat band, i.e. $N_e=6N_c$ ($\theta=0$) and $2N_c$ ($\theta=\theta_{\rm SFB}$) for pyrochlore lattice systems. 
In Figs.~\ref{f4}(a) and (b), $P_\sigma (\vb*{k}, \omega)$ in Eq.~(\ref{eq:spectral_function}) is shown for $U/t_{\rm eff}=1$, where there is a constant shift for half of the bands caused by Hartree-Fock approximation. 
Although FB-FM and FB-CS are interrelated by $\mathcal{U}_{\Gamma}$ and the peak positions are equivalent, the distributions of weights totally differ. 
The FB-FM state has large weights on the upper half of the dispersive band, whereas FB-CS state has weights on the lower half, including the flat band. 
\par
Let us next focus on the spectral weight of the lowest flat band, ${\cal S}_\sigma(\vb*{k})=P_\sigma(\vb*{k},\omega=0)$. 
Here, we extend Eq.~(\ref{eq:sk}) to the case where the polarization vector of the added particle's spin is denoted by $\vb*{e}$, namely, $a^\dagger_{\vb*{k}\vb*{e}}=\sum_\sigma\braket*{\sigma}{\vb*{e}}a^\dagger_{\vb*{k}\sigma}$ where $\ket*{\vb*{e}}=(\cos(\theta_{\rm e}/2), \sin(\theta_{\rm e}/2)e^{i\phi_{\rm e}})^{\rm T}$. 
The spectral weight is determined not solely by the relative angles between the polarization vectors $\vb*{p}$ and $\vb*{e}$. 
The complexity arises from the combination of these angles with the rotation axes of the local SU(2) gauge transformation.
To clarify the dependence on these vectors and the difference between FB-FM and FB-CS, 
we decompose the structure factor into a component ${\cal S}_{\mu\nu}(\vb*{k})$ common to all 
and its prefactor $v_{\mu\nu}$ describing these details; 
\begin{eqnarray}
\label{eq:Sek}&& {\cal S}_{\vb*{e}}(\vb*{k})=\sum_{\mu\nu} v_{\mu\nu} {\cal S}_{\mu\nu}(\vb*{k}), \\
\label{eq:Smunuk}&& {\cal S}_{\mu\nu}(\vb*{k}) = \frac{1}{n_s}\sum_{n \in {\rm FB}} e^{i\vb*{k}\cdot(\vb*{r}_\mu-\vb*{r}_\nu)}[\vb*{v}_n(\vb*{k})]^*_\mu [\vb*{v}_n(\vb*{k})]_\nu, 
\end{eqnarray}
where we use the Bloch state $\vb*{v}_n(\vb*{k})$ in Eq.~(\ref{eq:fbblchwf}). 
The vertex function $v_{\mu\nu}$ is represented as 
\begin{equation}\label{eq:vertexfunction2}
v_{\mu\nu} = \begin{dcases}
|\braket*{\vb*{e}}{\vb*{p}}|^2 & (\theta=0), \\
\mel*{\vb*{p}}{\Gamma^\dagger_\mu}{\vb*{e}} \mel*{\vb*{e}}{\Gamma_\nu}{\vb*{p}} & (\theta=\theta_{\rm SFB}).
\end{dcases}
\end{equation}
Since ${\cal S}^*_{\mu\nu}(\vb*{k})={\cal S}_{\nu\mu}(\vb*{k})$ and $v^*_{\mu\nu}=v_{\nu\mu}$, the structure factor is rewritten as
\begin{equation}
{\cal S}_{\vb*{e}}(\vb*{k})=\sum_{\mu\nu}  \Re \big[ v_{\mu\nu} {\cal S}_{\mu\nu}(\vb*{k}) \big]. 
\end{equation}
When $\theta=\theta_{\rm SFB}$, the vertex function is expressed as
\begin{equation}\label{eq:vertexfunction}
\begin{split}
v_{\mu\nu} &= \frac{1}{2} \big[ (1-\vb*{e}\cdot\vb*{p}) (\hat{\vb*{m}}_\mu\cdot\hat{\vb*{m}}_\nu) \\
&\qquad+(\hat{\vb*{m}}_\mu\cdot\vb*{e})(\hat{\vb*{m}}_\nu\cdot\vb*{p}) +(\hat{\vb*{m}}_\mu\cdot\vb*{p})(\hat{\vb*{m}}_\nu\cdot\vb*{e}) \\
&\qquad+i(\vb*{p}-\vb*{e})\cdot(\hat{\vb*{m}}_\mu\cross\hat{\vb*{m}}_\nu) \big], 
\end{split}
\end{equation}
whose detail is shown in Appendix~\ref{app:derivation}. 
This formula represents the complex behavior of the FB-CS spectrum, 
in contrast to the FB-FM state with $v_{\mu\nu}=(1+\vb*{e}\cdot\vb*{p})/2$ that depends solely on 
the relative angle between $\vb*{e}$ and $\vb*{p}$. 
\par
Figure \ref{f4}(c) shows the structure factor ${\cal S}_{\vb*{e}}(\vb*{k})$ 
of a pyrochlore lattice flat-band state for various polarization vectors. 
The band touching points exhibit the pinch-point singularity. 
In the FB-CS state, the intensity distribution varies depending on $\theta_{\rm e}$,  
while in the FB-FM state the intensity changes, but not its shape. 
\par
If the measurement is not spin-resolved, the structure factor is given by
\begin{equation}
\begin{split}
{\cal S}(\vb*{k}) &=\sum_{\pm\vb*{e}} {\cal S}_{\vb*{e}}(\vb*{k}) \\
&= \sum_{\mu\nu}  \Re \big[ (\hat{\vb*{m}}_\mu\cdot\hat{\vb*{m}}_\nu+i\vb*{p}\cdot(\hat{\vb*{m}}_\mu\cross\hat{\vb*{m}}_\nu) ) {\cal S}_{\mu\nu}(\vb*{k}) \big]. 
\end{split}
\label{eq:skunpolar}
\end{equation}
To highlight the feature of Eq.(\ref{eq:skunpolar}), %%particular to ``spinor-ice metal", 
we refer to the magnetic structure factor of a conventional pyrochlore spin ice obtained for 
the large-$N$ limit given as 
\begin{equation}
{\cal S}^{\rm SI}(\vb*{k}) = \sum_{\mu\nu} (\hat{\vb*{m}}_\mu\cdot\hat{\vb*{m}}_\nu) {\cal S}_{\mu\nu}(\vb*{k}), 
\label{eq:spinice}
\end{equation}
which is similar to the first term of Eq.(\ref{eq:skunpolar}). 
Indeed, when the second term of Eq.(\ref{eq:skunpolar}) is relevant, 
which occurs for example, in the system with broken site-centered inversion symmetry, 
the spectrum will be significantly modified by the additional spinor degrees of freedom 
of the flat-band wave function. 
We show in Appendix~\ref{app:inversionsymmetry} the proof that 
if the lattice has site-centered inversion symmetry, the structure factor ${\cal S}_{\mu\nu}(\vb*{k})$ does not have an imaginary part 
and reduces to 
\begin{equation}
{\cal S}(\vb*{k}) = \sum_{\mu\nu} (\hat{\vb*{m}}_\mu\cdot\hat{\vb*{m}}_\nu) {\cal S}_{\mu\nu}(\vb*{k}),
\end{equation}
which is identical to Eq.~(\ref{eq:spinice}). 
\par
Figure~\ref{f4}(d) shows the corresponding spin-unpolarized structure factor for 
the FB-FM and FB-CS states, in which the second term of Eq.(\ref{eq:skunpolar}) vanishes. 
Accordingly, they do not depend on the polarization of the ground state. 
Figure~\ref{f5} shows the counterexample, 
the hyperkagome lattice without site-centered inversion symmetry, 
which is the 1/4-depleted pyrochlore lattice and has $n_s=12$ sublattices in the unit cell (see Appendix~\ref{app:pyrochlore}), and Na$_4$Ir$_3$O$_8$ is known as a typical compound that takes structure~\cite{PhysRevLett.99.137207,PhysRevLett.115.047201,PhysRevLett.99.037201,PhysRevLett.100.227201,PhysRevLett.101.197201,MasafumiUdagawa_2009,PhysRevB.78.094403}. 
Taking the SU(2) gauge field $U_{ij}$ common to the pyrochlore lattice, 
its band structure has the SFB for $\theta_{\rm SFB}=-2\arctan2\sqrt{2}$ as shown in Fig.~\ref{f5}(b). 
The structure factor ${\cal S}(\vb*{k})=\sum_{\pm\vb*{e}} {\cal S}_{\vb*{e}}(\vb*{k})$ in the $(hhl)$ plane 
rotates its gradation intensity when $\theta_{\rm p}$ is varied, meaning that 
the second term of Eq.(\ref{eq:skunpolar}) is activated. 
%
%*%*%*%*%*%*%*%*%*%*%*%*%*%*%*%*%*%*%*%*%*%*%*%*%*%*%*%*%*
%*%*%*%*%*%*%*%*%*%*%*%*%*%*%*%*%*%*%*%*%*%*%*%*%*%*%*%*%*
\section{Correlated and nearly flat band states}
\label{sec:correlation}
%*%*%*%*%*%*%*%*%*%*%*%*%*%*%*%*%*%*%*%*%*%*%*%*%*%*%*%*%*
%*%    phase diagram
%%
We now focus on the pyrochlore lattice relevant to several $5d$ electron systems, 
and examine the sustainability of pinch point under the correlation and other realistic effects. 
We further see how they manifest in the experimentally measurable quantities. 

%*%*%*%*%*%*%*%*%*%*%*%*%*%*%*%*%*%*%*s
\subsection{Effect of inter-site Coulomb term}
Here, we examine how pinch points may deteriorate by the slight destruction of flat bands or the lifting of band touching. 
Two factors are examined: Introducing finite dispersion, i.e., allowing for nearly flat-band systems, by setting $\theta \ne \theta_{\rm SFB}$, and taking finite intersite Coulomb interaction $V > 0$ in addition to $U >0$. 
We set the electron number to 1/4-filling, $N_e=2N_c$. 
\par
Figure~\ref{f6}(a) shows a $\theta$-$V/t_{\rm eff}$ phase diagram for $U/t_{\rm eff}=5$ 
obtained by the Hartree-Fock approximation. 
The paramagnetic metallic phase (PM) appears for $\theta>\theta_c \sim -\pi/2$ with no symmetry breaking. 
For $\theta<\theta_c$, the flat band develops and suppresses the kinetic energy scale, 
which drives the symmetry breaking to the CS phase.  
In the CS phase, the scalar spin chirality 
$\chi_{123}=\ev*{\vb*{S}_1}\cdot(\ev*{\vb*{S}_2}\cross\ev*{\vb*{S}_3})$ develops as shown in Fig.~\ref{f6}(b). 
\par
For $V/t_{\rm eff}\gtrsim 3$, a 1-in-3-out charge ordered (CO) phase is observed, 
having one charge rich and three charge-poor sublattice sites with $\ev*{n_{\rm in}}\simeq2$, 
$\ev*{n_{\rm out}}\simeq0$. 
However, such a state is unphysical and is supported by the artificially large value of $V/U$. 
There is a more natural 3-in-1-out CO solution with $\ev*{n_{\rm in}}\simeq2/3$ and $\ev*{n_{\rm out}}\simeq0$, which has slightly higher energy than the 1-in-3-out state. 
Since the 3-in-1-out state is observed in the relevant material, 
$\rm CsW_2O_6$ \cite{Okamoto2020,Okamoto2024}, we will focus on this phase shortly. 
\par
%*%*%*%*%*%*%*%*%*%*%*%*%*%*%*%*%*%*%*%*%*%*%*%*%*%*%*%*%*
%*%    small V
We first examine the small $V$ case by varying the degree of SOC represented by $\theta$ [see Eq.(\ref{eq:Ham_kin2})]. 
In the CS phase, the integrated spectral weight of the occupied band corresponding to ${\cal S}(\vb*{k})=P(\vb*{k},\omega=0)$ can still exhibit a bowtie-like structure as a remnant of pinch point in a relatively wide range marked with shades in Fig.~\ref{f6}(c), 
where we show the width of the occupied band as a function of $\theta$. 
For $\theta\sim \theta_{\rm SFB}$, a nearly flat band is persistent. 
Figure~\ref{f6}(d) shows the bowtie-like ${\cal S}(\vb*{k})$ for $\theta=-5\pi/6$ where the bandwidth is 
$w/t_{\rm eff}=0.57$.
Whereas for $\theta=-2\pi/3$ with $w/t_{\rm eff}=1.46$ the structure fades out. 
We find the precursor of a half-moon; the spectrum has a larger weight opposite to the pinch point. 
This half-moon is carried by the curl-free component \cite{Yan2018} and is a signature that there is a pinch point in the nearby parameter region.%
%*%*%*%*%*%*%*%*%*%*%*%*%*%*%*%*%*%*%*%*%*%*%*%*%*%*%*%*%*
%*%    large V
%*%*%*%*%*%*%*%*%*%*%*%*%*%*%*%*%*%*%*%*%*%*%*%*%*%*%*%*%*
\par 
For large $V$, as mentioned above, we physically expect the 3-in-1-out CO state to occur. 
We take 16 sites per unit cell based on the experimental observation of a 3-in-1-out CO structure, 
where the electrons occupy the hyperkagome lattice relevant to CsW$_2$O$_6$ \cite{Okamoto2020}. 
Figure~\ref{f6}(e) shows the energy of three phases we found as a function of $\theta$, 
each of which has the same charge distribution, $\ev*{n_{\rm in}} \simeq 2/3$ and $\ev*{n_{\rm out}} \simeq 0$. 
The three phases are distinguished by bond order parameters [Fig.~\ref{f6}(e)] and are denoted by CO-I, CO-II, and CO-III. 
The CO-III phase exists as a local minimum only for $\theta_{\rm SFB}$ and does not disrupt the flat band. 
The other two exist as local energy minima over a wider parameter range. 
The integrated spectral intensities in the $(hk0)$ plane and band structures for $\theta_{\rm SFB}$ for the three phases 
are shown in Fig.~\ref{f6}(f). 
Although there is no remarkable difference in the band structures between the CO-I and CO-II phases, 
their spectral intensities differ. 
However, the singularity is completely wiped out. 
In the CO-III phase, pinch points are observed since the flat band is preserved. 
This phase is particular in that the CO pattern is not regular, as it allows various types 
of 3-in-1-out local correlations, which gives us a chance to have pinch points.  
%
%
%*%*%*%*%*%*%*%*%*%*%*%*%*%*%*%*%*%*%*%*%*%*%*%*%*%*%*%*%*
%*%*%*%*%*%*%*%*%*%*%*%*%*%*%*%*%*%*%*%*%*%*%*%*%*%*%*%*%*
\subsection{Relevance with experiments}
\label{sec:exp}
To take account of the realistic situation, we focus on the ARPES experiment 
and see how our spectral weight $P_\sigma (\vb*{k}, \omega)$ 
is related through the matrix element effect \cite{RMP_2003, Moser_2017, Day2019, Boschini2020,RMP_2021,Schuler2022}.  
The photoemission intensity is given as 
\begin{equation}
I(\vb*{k}, \omega) = I_0 (\vb*{k}, h\nu, \vb*{A}) f(\omega) A(\vb*{k}, \omega), 
\end{equation}
where $f(\omega)$ is the Fermi-Dirac distribution and $A(\vb*{k}, \omega)$ is the single-particle spectral function given by 
\begin{equation}
A(\vb*{k}, \omega) = -\frac{1}{\pi} \Tr \Im G^{R}_{\gamma\gamma'}(\vb*{k}, \omega). 
\end{equation}
Here, $G^{R}_{\gamma\gamma'}(\vb*{k}, \omega)$ is the retarded Green's function and $\gamma$ denotes the internal degrees of freedom of the electrons, such as spin, orbital, and sublattice. 
The prefactor $I_0 (\vb*{k}, h\nu, \vb*{A})$ contains the square of the matrix element effects $M_{fi}(\vb*{k})$, which is represented as 
\begin{equation}
M_{fi}(\vb*{k}) = \mel*{\phi_f(\vb*{k})}{\mathcal{H}_{\rm LM}}{\phi_i(\vb*{k})}, 
\end{equation}
within the sudden approximation. 
Here, $\mathcal{H}_{\rm LM}$ is the Hamiltonian describing the light-matter interaction, and $\ket*{\phi_i(\vb*{k})}$ and $\ket*{\phi_f(\vb*{k})}$ are the initial and final one-electron states, respectively, in the photoemission process. 
In general, the spectral function $A(\vb*{k}, \omega)$ exhibits several peaks indicating the excited levels, whereas the intensity of ARPES spectrum is modulated by the matrix element effects, which contains microscopic information about the initial state; they include the orbital character of the Bloch electron \cite{Zhang2012, Cao2013, King2014, Watson2015, Matt2018} and the relative phases between sublattices \cite{Mucha2008, Gierz2011, Liu2011, Zhu2013}. 
\par
For simplicity, we assume that the final state $\ket*{\phi_f(\vb*{k})}$ is a plane wave $\ket*{\vb*{k}_f}$ with momentum $\hbar\vb*{k}_f$ and the initial state $\ket*{\phi_n(\vb*{k})}$ with band index $n$ is expanded by the tight-binding basis as
\begin{equation}
\ket*{\phi_n(\vb*{k})} \propto \sum_l e^{i\vb*{k}\cdot\vb*{R}_l} \sum_{\mu l_z \sigma} C_{n;\mu l_z \sigma}(\vb*{k}) \ket*{\vb*{R}_l +\vb*{r}_\mu, l_z, \sigma}, 
\end{equation}
where $l_z$ is the magnetic quantum number. 
Then, $M_{fi}(\vb*{k})$ is described by the polarization term, orbital term, and momentum conservation term \cite{Moser_2017}. 
The polarization term represents the effect of the geometry of the experiment and the polarization of the photons, and the orbital term depends on the microscopic structure of the initial state. 
We focus here only on the orbital term. 
In this case, $M_{fi}(\vb*{k})$ is given by 
\begin{equation}
M_{fi}(\vb*{k}) \propto \sum_{\mu l_z \sigma} C_{n;\mu l_z \sigma}(\vb*{k}) \braket*{\vb*{k}_f}{l_z, \sigma}. 
\end{equation}

%*%*%*%*%*%*%*%*%*%*%*%*%*%*%*%*%*%*%*%*%*%*%*%*%*%*%*%*%*
\begin{figure}[t]
	\begin{center}
		\includegraphics[width=8.6cm]{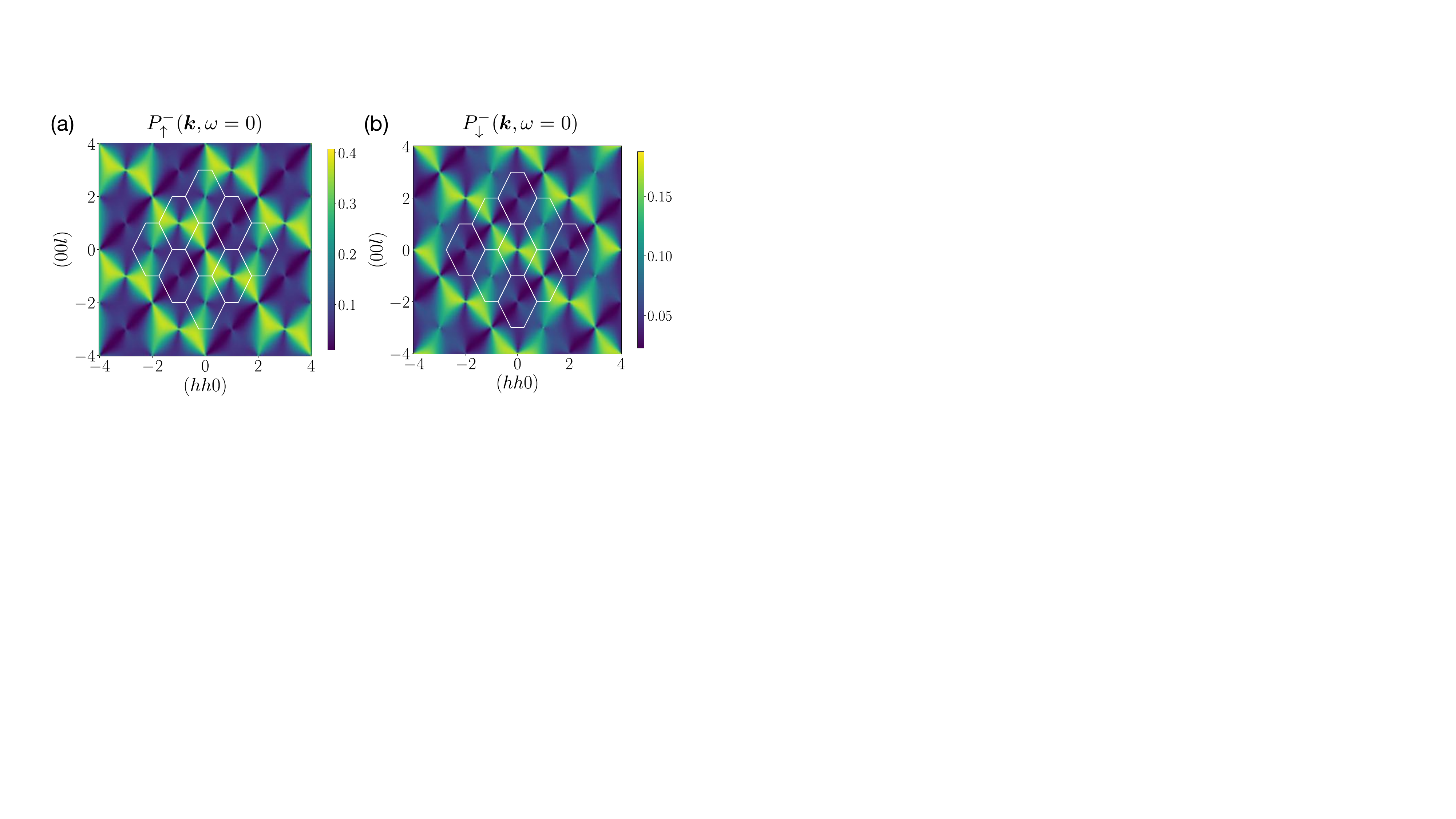}
		\caption{Spectral weight of the lowest flat band in the $(hhl)$ plane calculated for $\rm CsW_2O_6$. (a) $P_\uparrow(\vb*{k}, \omega=0)$ and (b) $P_\downarrow(\vb*{k}, \omega=0)$. The white lines represent the BZ boundaries.}
		\label{f7}
	\end{center}
\end{figure}
%*%*%*%*%*%*%*%*%*%*%*%*%*%*%*%*%*%*%*%*%*%*%*%*%*%*%*%*%*

Next, we relate this matrix element effect to our spectral weight. 
Considering a single-orbital model for simplicity, we find $C_{n;\mu l_z \sigma}(\vb*{k})=e^{-i\vb*{k}\cdot\vb*{r}_\mu}[\vb*{u}_{n}(\vb*{k})]_{\mu\sigma}$ and 
\begin{equation}
M_{fi; \sigma}(\vb*{k}) \propto \sum_{\mu} e^{-i\vb*{k}\cdot\vb*{r}_\mu}[\vb*{u}_{n}(\vb*{k})]_{\mu\sigma} \braket*{\vb*{k}_f}{l_z}, 
\end{equation}
assuming a spin selectivity in the photoemission process. 
Then, the photoemission intensity is given by 
\begin{equation}
|M_{fi; \sigma}(\vb*{k})|^2 \propto |\braket*{\vb*{k}_f}{l_z}|^2 \Big|\sum_{\mu} e^{-i\vb*{k}\cdot\vb*{r}_\mu} [\vb*{u}_{n}(\vb*{k})]_{\mu\sigma} \Big|^2, 
\end{equation}
which includes our spectral weight. 
However, we need to be careful when applying it to the pyrochlore systems with strong SOC. 
In $4d$ and $5d$ pyrochlore compounds, the $t_{2g}$ orbitals are split by the crystal field and strong SOC, giving rise to Kramers doublets, which can be regarded as pseudospins \cite{Khomskii2016}. 
When we focus on one of the pseudospins, the Hamiltonian is expressed as Eq.~(\ref{eq:Ham_kin}), 
where the spin index $\sigma$ is regarded as that of the pseudospin $\Lambda=\Uparrow, \Downarrow$. 
Furthermore, if we take into account the anisotropy of the quantization axis owing to the octahedron formed by the ligand ions, the representation of the mixture of orbitals and spins in the pseudospin in the {\it global} axis depends on the sublattice; 
\begin{equation}
\ket*{\Lambda_\mu} = \sum_{l_z, \sigma} X^{\mu\Lambda_\mu}_{l_z, \sigma} \ket*{l_z, \sigma}. 
\end{equation}
Then, we find 
\begin{equation}
C_{n;\mu l_z \sigma}(\vb*{k}) = \sum_{\Lambda_\mu} X^{\mu\Lambda_\mu}_{l_z, \sigma} e^{-i\vb*{k}\cdot\vb*{r}_\mu} [\vb*{u}_n(\vb*{k})]_{\mu\Lambda_\mu}, 
\end{equation}
where $\vb*{u}_n(\vb*{k})$ is obtained by diagonalizing the Bloch Hamiltonian. 
The photoemission intensity with spin selectivity is given by 
\begin{equation}\label{eq:ARPES1}
|M_{fi; \sigma}(\vb*{k})|^2 \propto \Bigg| \sum_{\mu\Lambda_\mu} \tilde{X}^{\mu\Lambda_\mu}_{\vb*{k}_f} e^{-i\vb*{k}\cdot\vb*{r}_\mu}[\vb*{u}_n(\vb*{k})]_{\mu\Lambda_\mu} \Bigg|^2, 
\end{equation}
where
\begin{equation}
\tilde{X}^{\mu\Lambda_\mu}_{\vb*{k}_f} = \sum_{l_z} X^{\mu\Lambda_\mu}_{l_z, \sigma} \braket*{\vb*{k}_f}{l_z}. 
\end{equation}
On the other hand, our spectral weight is rewritten as
\begin{equation}\label{eq:ARPES2}
\begin{split}
|\mel*{\Psi^{N\pm 1}_{\vb*{k}n}} {c^{\pm}_{\vb*{k}\sigma}}{\Psi^{N}_g} |^2 &= | \mel*{\Psi^{N}_g}{\{ c^{\mp}_{n\vb*{k}}, a^{\pm}_{\vb*{k}\sigma} \} }{\Psi^{N}_g} |^2 \\
&\simeq \Bigg| \sum_{\mu\Lambda_\mu} \Big( \sum_{l_z} X^{\mu\Lambda_\mu}_{l_z, \sigma} \Big) e^{-i\vb*{k}\cdot\vb*{r}_\mu}[\vb*{u}_n(\vb*{k})]_{\mu\Lambda_\mu} \Bigg|^2, 
\end{split}
\end{equation}
since 
\begin{equation}
c^\dagger_{n\vb*{k}} = \sum_{\mu\Lambda_\mu l_z \sigma} X^{\mu\Lambda_\mu}_{l_z, \sigma} [\vb*{u}_n(\vb*{k})]_{\mu\Lambda_\mu} c^\dagger_{\vb*{k}\mu l_z \sigma}. 
\end{equation}
Here, we approximate as $\mel*{\Psi^{N}_g}{\{ c^{\mp}_{\vb*{k}\mu l_z \sigma}, c^{\pm}_{\vb*{k}\mu'\sigma} \} }{\Psi^{N}_g} \simeq \delta_{\mu\mu'}$, neglecting the orbital-dependent difference in this factor. 
Comparing Eqs.~(\ref{eq:ARPES1}) and (\ref{eq:ARPES2}), we find that these are identical except for the factor $\braket*{\vb*{k}_f}{l_z}$. 
Figure~\ref{f7} shows the spectral weights with $\sigma=\uparrow, \downarrow$ for $\rm CsW_2O_6$. 
The derivation of the coefficients $\{X^{\mu\Lambda_\mu}_{l_z, \sigma}\}$ is provided in Appendix~\ref{microscopic_derivation}. 
Although different from Fig.~\ref{f3}, the singularity at the band touching point is still observed.

%*%*%*%*%*%*%*%*%*%*%*%*%*%*%*%*%*%*%*%*%*%*%*%*%*%*%*%*%*%*%*%*%*%*%*%*%*%*%*%*%*%*%*%*%*%*%*%*%*%*%*%*%*%*%*%*%*%*
%*%*%*%*%*%*%*%*%*%*%*%*%*%*%*%*%*%*%*%*%*%*%*%*%*%*%*%*%*%*%*%*%*%*%*%*%*%*%*%*%*%*%*%*%*%*%*%*%*%*%*%*%*%*%*%*%*%*
\section{Summary and Discussion}
\label{sec:summary}
We investigated the excitation spectrum of the electronic systems on kagome and pyrochlore lattices 
with spin-orbit coupling (SOC) as well as Coulomb interactions. 
The main focus was to clarify how the electronic spin degrees of freedom act on the 
pinch-point singularity that manifests in the spectral function of a flat band realized in these systems. 
\par
Pinch points have been widely studied as experimental signatures of classical spin liquids, 
particularly in spin ice state of pyrochlore insulating magnets. 
There the low-energy excitations from a flat band are effectively described by bosons, 
and the singularity at the band-touching point $\vb*{k}^*$ means that 
the flat-band state at this point is described only by the noncontractible loop states (NLS),  
which form one of the two major classes of flat-band eigenstates. 
The other class consists of compact localized states (CLS), 
closed loops of finite length, responsible for $\vb*{k}$ off $\vb*{k}^*$. 
Importantly, the CLS and NLS are topologically distinct (cannot be adiabatically connected to each other) 
and so are the general $\vb*{k}$ points and $\vb*{k}^*$. 
Previously, we analytically demonstrated the CLS-NLS correspondence when approaching the 
singular point, $\vb*{k}\rightarrow\vb*{k}^*$, 
and showed how NLS spanning different bond directions interfere to produce 
an angular-dependent gradation of the pinch-point spectrum \cite{Udagawa2024}. 
The formulation developed in Ref.[\onlinecite{Udagawa2024}] can explain in other way round the behavior of the spectrum 
we presented in this paper. 
\par
In this work, we explored how spin degrees of freedom of electrons 
influence the spectrum, when SOC introduces chirality into the flat-band ground state. 
Such extra degrees of freedom were not present in the bosonic excitation (magnon that is the single degrees of freedom) 
of the classical magnet. 
When spins in the flat band are ``polarized" in a direction $\bm{p}$ owing to SOC, the vertex function 
acting as a prefactor to the spectral function depends on the relative angles between $\bm{p}$, 
the spin polarization vector $\bm{e}$ of the injected electron, 
and the SU(2) gauge transformation vector $\hat{\vb*{m}}_\mu$, which is determined by the SOC strength. 
If a spin-unpolarized electron ($\bm{e} = 0$) is introduced and the system maintains site-centered inversion symmetry, 
the spectrum reduces to the standard spin-ice pinch point. 
Otherwise, the pinch point can rotate with changes in $\bm{p}$ or vary in intensity, 
suggesting that SOC flat bands could be experimentally probed by rotating $\bm{e}$ to infer internal magnetic structures.
\par
In realistic materials, achieving a perfectly flat band through fine-tuning of material parameters is impractical, 
and additional effects such as Coulomb interactions may obscure key spectral features. 
Thus, another focus of our study is the robustness of the pinch-point singularity and its transformation under perturbations.
By incorporating substantial Coulomb interactions and varying band parameters away from the perfect flat-band limit, 
we find that, although the singularity is lifted because of band-touching removal, 
remnants of pinch-point features persist as long as the system remains in the chiral spin state. 
However, in a transition to a conventional charge-ordered phase, the singularity is completely wiped out. 
Additionally, we provided realistic insights into matrix-element effects 
in angle-resolved photoemission spectroscopy and applied our findings to CsW$_2$O$_6$, 
where Kramers doublets carrying pseudospins, a combination of orbital and spin momentum, play a crucial role.

%*%*%*%*%*%*%*%*%*%*%*%*%*%*%*%*%*%*%*%*%*%*%*%*%*%*%*%*%*
%*%*%*%*%*%*%*%*%*%*%*%*%*%*%*%*%*%*%*%*%*%*%*%*%*%*%*%*%*
\begin{acknowledgments}
This work is supported by KAKENHI Grants No. JP20H05655, JP21H05191, JP21K03440, JP22H01147, JP23KJ0783, and JP23K22418
from JSPS of Japan.
\end{acknowledgments}

%*%*%*%*%*%*%*%*%*%*%*%*%*%*%*%*%*%*%*%*%*%*%*%*%*%*%*%*%*
\appendix
\section{Conventions of lattices}\label{app:pyrochlore}
In this appendix, we list the lattice conventions for pyrochlore, kagome, and hyperkagome lattices.

\subsection{Pyrochlore lattice}
The pyrochlore lattice consists of four sublattices in a unit cell with the lattice vectors given by $\vb*{a}_1=a(1,-1,0)/2$, $\vb*{a}_2=a(0,-1,1)/2$, and $\vb*{a}_3=a(1,0,1)/2$, where $a$ is a lattice constant. 
The corresponding reciprocal lattice vectors are given by $\vb*{b}_1=2\pi(1,-1,-1)/a$, $\vb*{b}_2=2\pi(-1,-1,1)/a$, and $\vb*{b}_3=2\pi(1,1,1)/a$. The $(hhl)$ plane shown is spanned by $\vb*{b}_2$ and $\vb*{b}_3$. 
The coordinates of four sublattices in the unit cell are given by $\vb*{r}_0=(0,0,0)$, $\vb*{r}_1=\vb*{a}_1/2=a(1,-1,0)/4$, $\vb*{r}_2=\vb*{a}_2/2=a(0,-1,1)/4$, and $\vb*{r}_3=\vb*{a}_3/2=a(1,0,1)/4$. 
The highly symmetric points in the first BZ shown in Fig.~\ref{f2}(b) are $\Gamma=(0,0,0)$, ${\rm X}=(0,0,2\pi)/a$, ${\rm K}=(3\pi/2,0,3\pi/2)/a$, ${\rm W}=(\pi,0,2\pi)/a$, and ${\rm L}=(\pi,\pi,\pi)/a$. 

\subsection{Kagome lattice}
The lattice convention of kagome lattice is 
$\vb*{a}_1=a(1,0)$ and $\vb*{a}_2=a(1/2, \sqrt{3}_2)$ 
and  $\vb*{b}_1=2\pi(1,-1/\sqrt{3})/a$ and $\vb*{b}_2=2\pi(0, 2/\sqrt{3})/a$. 
The sublattice coordinates for $n_s=3$ are given by 
$\vb*{r}_0=(0,0)$, $\vb*{r}_1=\vb*{a}_1/2=a(1/2,0)$, and $\vb*{r}_2=\vb*{a}_2/2=a(1/4,\sqrt{3}/4)$. 
The highly symmetric points in the first BZ shown in Fig.\ref{f2} are $\Gamma=(0,0)$, ${\rm K}=(4\pi/3, 0)/a$, and ${\rm M}=(\pi, \pi/\sqrt{3})/a$. 

\subsection{Hyperkagome lattice}
The hyperkagome lattice shown in Fig.~\ref{f5}(a) is a 1/4-depleted pyrochlore lattice. 
We have $n_s=12$ sublattices and we take 
$\vb*{a}_1=a(1,0,0)$, $\vb*{a}_2=a(0,1,0)$, and $\vb*{a}_3=a(0,0,1)$, 
$\vb*{b}_1=2\pi(1,0,0)/a$, $\vb*{b}_2=2\pi(0,1,0)/a$, and $\vb*{b}_3=2\pi(0,0,1)/a$. 
The 12 lattice sites are obtained from the 16 sites of the pyrochlore lattice, 
$\{(l_i,\mu)\ |\ l_i, \mu=1,2,3,4\}$ with $\vb*{R}_{l_2}=\vb*{R}_{l_1}-\vb*{a}_2+\vb*{a}_3$, $\vb*{R}_{l_3}=\vb*{R}_{l_1}-\vb*{a}_1+\vb*{a}_3$, and $\vb*{R}_{l_4}=\vb*{R}_{l_1}+\vb*{a}_3$, 
removing four of them, $(l_1, 4)$, $(l_2, 1)$, $(l_3, 3)$, and $(l_4, 2)$. 
%*%*%*%*%*%*%*%*%*%*%*%*%*%*%*%*%*%*%*%*%*%*%*%*%*%*%*%*%*
%*%*%*%*%*%*%*%*%*%*%*%*%*%*%*%*%*%*%*%*%*%*%*%*%*%*%*%*%*
\section{Derivation of Eq.~(\ref{eq:vertexfunction})}\label{app:derivation}
In this appendix, we get Eq.~(\ref{eq:vertexfunction}) from Eq.~(\ref{eq:vertexfunction2}). 
Since the projector is expressed as $\ketbra*{\vb*{e}}{\vb*{e}}=(1+\vb*{e}\cdot\vb*{\sigma})/2$ and the local gauge transformation is $\Gamma_\mu=-i\hat{\vb*{m}}_\mu\cdot\vb*{\sigma}$, the vertex function $v_{\mu\nu}$ is given by
\begin{equation}
\begin{split}
v_{\mu\nu} &= \frac{1}{2} \mel*{\vb*{p}}{\Gamma^\dagger_\mu(1+\vb*{e}\cdot\vb*{\sigma})\Gamma_\nu}{\vb*{p}} \\
&= \frac{1}{2} \big[ \mel*{\vb*{p}}{(\hat{\vb*{m}}_\mu\cdot\vb*{\sigma})(\hat{\vb*{m}}_\nu\cdot\vb*{\sigma})}{\vb*{p}} \\
&\qquad +\mel*{\vb*{p}}{(\hat{\vb*{m}}_\mu\cdot\vb*{\sigma})(\vb*{e}\cdot\vb*{\sigma})(\hat{\vb*{m}}_\nu\cdot\vb*{\sigma})}{\vb*{p}}\big].
\end{split}
\end{equation}
Using these formulas,
\begin{eqnarray}
&& (\hat{\vb*{m}}_\mu\cdot\vb*{\sigma})(\hat{\vb*{m}}_\nu\cdot\vb*{\sigma}) \nonumber\\
&&= \hat{\vb*{m}}_\mu\cdot\hat{\vb*{m}}_\nu +i(\hat{\vb*{m}}_\mu\cross\hat{\vb*{m}}_\nu)\cdot\vb*{\sigma}, \\
&& (\hat{\vb*{m}}_\mu\cdot\vb*{\sigma})(\vb*{e}\cdot\vb*{\sigma})(\hat{\vb*{m}}_\nu\cdot\vb*{\sigma}) \nonumber\\
&&= (\hat{\vb*{m}}_\mu\cdot\vb*{e})(\hat{\vb*{m}}_\nu\cdot\vb*{\sigma}) +(\hat{\vb*{m}}_\mu\cdot\vb*{\sigma})(\hat{\vb*{m}}_\nu\cdot\vb*{e}) \nonumber \\
&&\quad -(\hat{\vb*{m}}_\mu\cdot\hat{\vb*{m}}_\nu)(\vb*{e}\cdot\vb*{\sigma})-i(\hat{\vb*{m}}_\mu\cross\hat{\vb*{m}}_\nu)\cdot\vb*{e},
\end{eqnarray}
and $\mel*{\vb*{p}}{(\vb*{n}\cdot\vb*{\sigma})}{\vb*{p}}=\vb*{n}\cdot\vb*{p}$, we obtain the desired Eq.~(\ref{eq:vertexfunction}). 

\section{Role of the site-centered inversion symmetry}\label{app:inversionsymmetry}
In this appendix, we prove that ${\cal S}_{\mu\nu}(\vb*{k})$ in Eq.~(\ref{eq:Smunuk}) for the site-centered inversion symmetric systems does not have the imaginary part; ${\cal S}^*_{\mu\nu}(\vb*{k})={\cal S}_{\mu\nu}(\vb*{k})$. 
To show this, we introduce a different convention for Fourier transformation than in the main text. 
\subsection{Bloch Hamiltonians}
In the main text, we introduce the Fourier transformation as
\begin{equation}
c^\dagger_{\vb*{k}\mu\sigma} = \frac{1}{\sqrt{N_c}} \sum_{l=0}^{N_c-1} e^{i\vb*{k}\cdot\vb*{R}_l} c^\dagger_{l\mu\sigma}, 
\end{equation}
and the corresponding Bloch Hamiltonian and its eigenvector have periodicity with respect to the reciprocal vector $\vb*{G}$; $\mathcal{H}(\vb*{k}+\vb*{G})=\mathcal{H}(\vb*{k})$ and $\vb*{u}_n(\vb*{k}+\vb*{G})=\vb*{u}_n(\vb*{k})$ with $\mathcal{H}(\vb*{k})\vb*{u}_n(\vb*{k})=\varepsilon_n(\vb*{k})\vb*{u}_n(\vb*{k})$. 
\par
In an alternative manner of Fourier transformation, 
\begin{equation}
c'^\dagger_{\vb*{k}\mu\sigma} = \frac{1}{\sqrt{N_c}} \sum_{l=0}^{N_c-1} e^{i\vb*{k}\cdot(\vb*{R}_l+\vb*{r}_\mu)} c^\dagger_{l\mu\sigma}, 
\end{equation}
the corresponding Bloch Hamiltonian and its eigenvector do not have periodicity with respect to the reciprocal vector $\vb*{G}$;
\begin{eqnarray}
&& \mathcal{H}'(\vb*{k}+\vb*{G}) = V(\vb*{G}) \mathcal{H}'(\vb*{k}) V(\vb*{G})^\dagger, \\
&& \vb*{u}'_n(\vb*{k}+\vb*{G}) = V(\vb*{G}) \vb*{u}'_n(\vb*{k}),
\end{eqnarray}
where $[V(\vb*{k})]_{\mu\sigma,\mu'\sigma'}=e^{-i\vb*{k}\cdot\vb*{r}_\mu}\delta_{\mu\mu'}\delta_{\sigma\sigma'}$. 
\par
These are connected by the unitary operator $V(\vb*{k})$ as
\begin{eqnarray}
&& \mathcal{H}'(\vb*{k}) = V(\vb*{k}) \mathcal{H}(\vb*{k}) V(\vb*{k})^\dagger, \\
&& \vb*{u}'_n(\vb*{k}) = V(\vb*{k}) \vb*{u}_n(\vb*{k}). 
\end{eqnarray}

\subsection{Structure factor for the site-centered inversion symmetric systems}
Following the convention of the Fourier transformation introduced above, ${\cal S}_{\mu\nu}(\vb*{k})$ is represented as 
\begin{equation}
{\cal S}_{\mu\nu}(\vb*{k}) = \frac{1}{n_s}\sum_{n \in {\rm FB}} [\vb*{u}'_n(\vb*{k})]^*_\mu [\vb*{u}'_n(\vb*{k})]_\nu.
\end{equation}
Since Bloch Hamiltonian $\mathcal{H}'(\vb*{k})$ is a real symmetric matrix in the site-centered inversion symmetric systems, $\vb*{u}'_n(\vb*{k})$ can be taken as a real vector. 
Therefore, ${\cal S}_{\mu\nu}(\vb*{k})$ is a real number.

%*%*%*%*%*%*%*%*%*%*%*%*%*%*%*%*%*%*%*%*%*%*%*%*%*%*%*%*%*
\begin{figure}[t]
	\begin{center}
		\includegraphics[width=8.6cm]{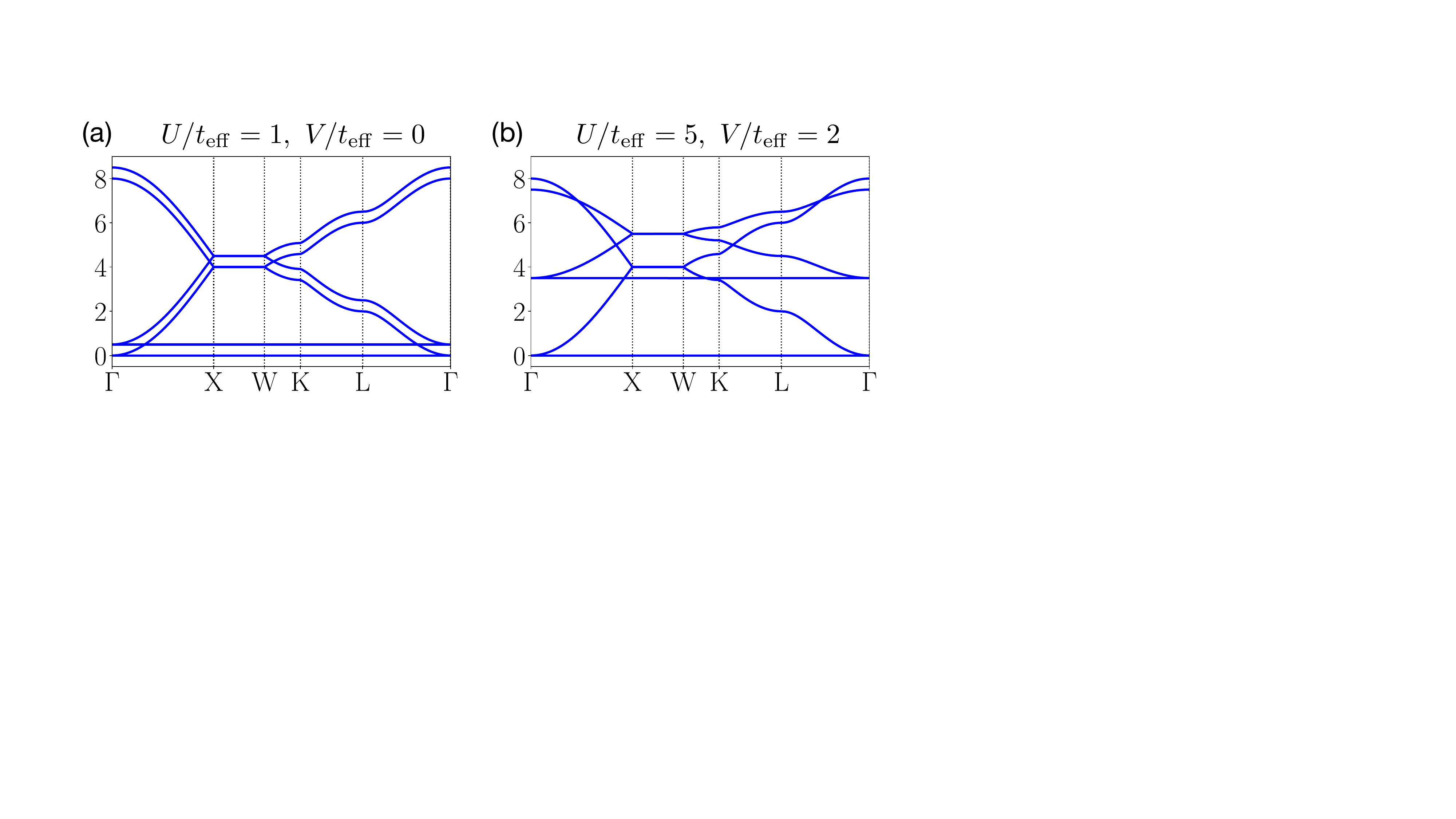}
		\caption{Band structures obtained by the Hartree-Fock approximation to the FB-FM (which is equal to FB-CS). 
			(a) $U/t_{\rm eff}=1$ and $V/t_{\rm eff}=0$. 
			(b) $U/t_{\rm eff}=5$ and $V/t_{\rm eff}=2$. }
		\label{fS8}
	\end{center}
\end{figure}
%*%*%*%*%*%*%*%*%*%*%*%*%*%*%*%*%*%*%*%*%*%*%*%*%*%*%*%*%*
\section{Mean-field Calculation}\label{app:mfcalculation}
We employ a Hartree-Fock approximation to Eq.~(\ref{eq:Ham_int}) and obtain 
\begin{widetext}
	\begin{equation}\label{eq:Ham_MF}
	\begin{split}
	\mathcal{H}^{\rm MF}_{\rm int} &= U\sum_i \Big( n_{i\uparrow} \ev*{n_{i\downarrow}}+\ev*{n_{i\uparrow}} n_{i\downarrow} - S_{i+}\ev*{S_{i-}} -\ev*{S_{i+}}S_{i-} -\det \chi_i \Big) \\
	&\quad +V\sum_{\langle i,j \rangle} \Big( n_{i} \expval*{n_{j}} +\expval*{n_{i}} n_{j} -\sum_{\sigma\tau}\big(c^\dagger_{i\sigma} c_{j\tau} \expval*{c^\dagger_{j\tau} c_{i\sigma}} + \expval*{c^\dagger_{i\sigma} c_{j\tau}} c^\dagger_{j\tau} c_{i\sigma}\big) -( \tr \chi_i) (\tr \chi_j) +\tr \*[\eta_{ij}\eta^\dagger_{ij}\*] \Big) , 
	\end{split}
	\end{equation}
\end{widetext}
where $\chi_i$ and $\eta_{ij}$ are the mean fields at site $i$ and the nearest neighbor pair $\langle i,j\rangle$ given by
\begin{eqnarray}
\label{eq:MF_chi}
&&\chi_i = \mqty( \ev*{c^\dagger_{i\uparrow}c_{i\uparrow}} & \ev*{c^\dagger_{i\uparrow}c_{i\downarrow}} \\ \ev*{c^\dagger_{i\downarrow}c_{i\uparrow}} & \ev*{c^\dagger_{i\downarrow}c_{i\downarrow}}) = \mqty( \ev*{n_{i\uparrow}} & \ev*{S_{i+}} \\ \ev*{S_{i-}} & \ev*{n_{i\downarrow}} ), \\
\label{eq:MF_eta}
&&\eta_{ij} = \mqty( \ev*{c^\dagger_{i\uparrow}c_{j\uparrow}} & \ev*{c^\dagger_{i\uparrow}c_{j\downarrow}} \\ \ev*{c^\dagger_{i\downarrow}c_{j\uparrow}} & \ev*{c^\dagger_{i\downarrow}c_{j\downarrow}}). 
\end{eqnarray}
The mean field for the FB-FM state (\ref{eq:gs_wf}) is given by
\begin{equation}
\chi_i = \frac{1}{2}\mqty(1 & 0 \\ 0 & 0), \quad \eta_{ij} = -\frac{1}{6} \mqty(1 & 0 \\ 0 & 0), 
\end{equation}
and for the FB-CS state (\ref{eq:gs_wf}) is given by
\begin{align}
\label{eq:mean_field_FM}
&\chi_1 = \frac{1}{6}\mqty( 1 & 1-i \\ 1+i & 2 ), \quad \chi_2 = \frac{1}{6}\mqty( 1 & -1+i \\ -1-i & 2 ), \nonumber\\
&\chi_3 = \frac{1}{6}\mqty( 1 & -1-i \\ -1+i & 2 ), \quad \chi_4 = \frac{1}{6}\mqty( 1 & 1+i \\ 1-i & 2 ), 
\nonumber\\
%\end{align}
%\begin{align}
&\eta_{12} = \frac{1}{18}\mqty( -1 & 1-i \\ -1-i & 2 ), \quad \eta_{13} = \frac{1}{18}\mqty( 1 & -1-i \\ 1+i & -2i ), \nonumber\\
&\eta_{14} = \frac{1}{18}\mqty( 1 & 1+i \\ 1+i & 2i ), \quad \eta_{23} = \frac{1}{18}\mqty( 1 & -1-i \\ -1-i & 2i ), \nonumber\\
&\eta_{24} = \frac{1}{18}\mqty( 1 & 1+i \\ -1-i & -2i ), \quad \eta_{34} = \frac{1}{18}\mqty( -1 & -1-i \\ 1-i & 2 ). 
\end{align}
As shown in Fig.~\ref{fS8}, the band structure in this mean-field potential for the FB-FM and FB-CS states 
are the same except for making the former upside down to get the latter, 
and the perfect flat bands remain. 
Their ground-state energy per site is given by
\begin{equation}
E_{\rm FM} = E_{\rm CS} = -t_{\rm eff} +\frac{2}{3} V. 
\end{equation}
The mean field for the 3-in-1-out CO-III state also keeps the flat-band structure 
for the same filling factor. 
Its ground-state energy per site is given by
\begin{equation}
E_{\rm CO} = -t_{\rm eff} +\frac{1}{12} U +\frac{7}{12} V. 
\end{equation}

%*%*%*%*%*%*%*%*%*%*%*%*%*%*%*%*%*%*%*%*%*%*%*%*%*%*%*%*%*
\begin{figure}[t]
	\begin{center}
	\includegraphics[width=7cm]{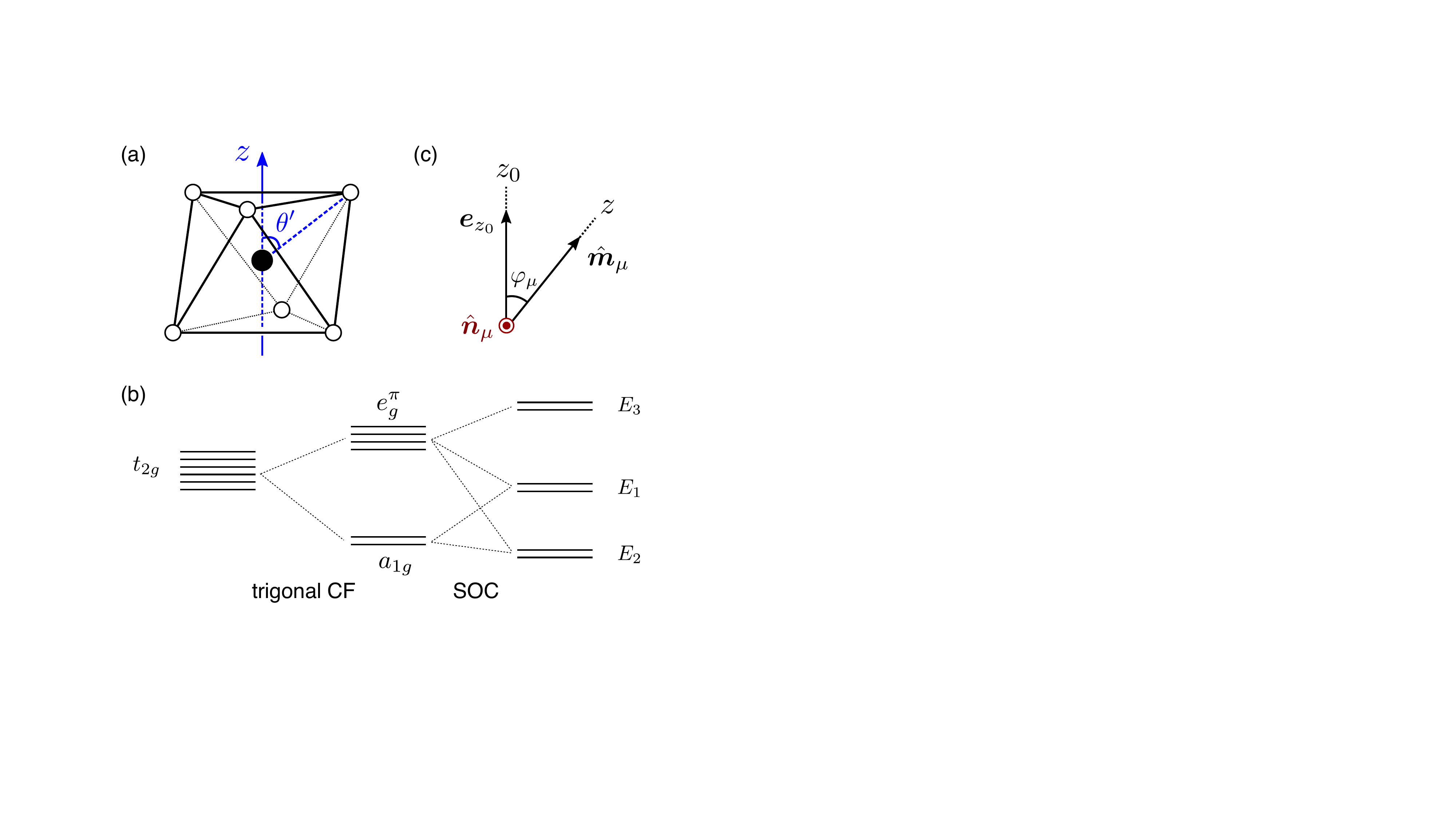}
\caption{(a) Trigonally distorted octahedron. Black and white circles denote metal and ligand ions, respectively. (b) Energy level splitting caused by trigonal CF and SOC. (c) Local rotation of angle $\varphi_\mu$ about the $\hat{\vb*{n}}_\mu$ axis to transform from the local quantization axis $z$ to the global quantization axis $z_0$.}
		\label{fS9}
	\end{center}
\end{figure}
%*%*%*%*%*%*%*%*%*%*%*%*%*%*%*%*%*%*%*%*%*%*%*%*%*%*%*%*%*

\section{Microscopic calculation of coefficients}\label{microscopic_derivation}
We microscopically derive the coefficients $\{X^{\mu\Lambda_\mu}_{l_z, \sigma}\}$ describing the mixture of orbitals and spins in the pseudospin for pyrochlore compounds. 
We start from a single transition metal ion surrounded by a trigonally distorted octahedron of ligand ions as shown 
in Fig.~\ref{fS9}(a). 
Each ligand ion carries a negative point charge $-Ze$ at position $\vb*{r}_i$ 
and the electrostatic energy of an electron sitting on the center ion at position $\vb*{r}$ 
is expressed as 
\begin{equation}
\mathcal{H}_{\rm CF}(\vb*{r}) = \sum_{i} \frac{Ze^2}{|\vb*{r}_i -\vb*{r}|}. 
\end{equation}
Since the $d$ orbitals at the center are well-localized, we can expand $\mathcal{H}_{\rm CF}$ using spherical harmonics as 
\begin{equation}\label{eq:CF}
\mathcal{H}_{\rm CF}(r, \theta, \varphi) = \sum_{k=0}^{\infty} \sum_{m=-k}^{k} r^k q_{km} C_{km}(\theta, \varphi), 
\end{equation}
where $q_{km}$ and $C_{km}(\theta, \varphi)$ are given as
\begin{align}
&q_{km} = \sqrt{\frac{4\pi}{2k+1}} \frac{Ze^2}{r_0} \sum_{i=1}^6 Y^*_{km}(\theta_i, \varphi_i), \\
&C_{km}(\theta, \varphi) = \sqrt{\frac{4\pi}{2k+1}} Y_{km}(\theta, \varphi). 
\end{align}
The octahedron has threefold rotational symmetry about the [111] direction set as the local $z$ axis. 
The degree of trigonal distortion is expressed by the angle $\theta'$ between $z$ axis and the direction to the ligand ion, and $\theta'= \theta_0=\arccos(1/\sqrt{3}) \simeq 54.74^\circ$ corresponds to an ideal octahedron. 
From the symmetry of the trigonally distorted octahedron, (i) threefold rotational symmetry, (ii) inversion symmetry, and (iii) mirror symmetry, the form of CF Hamiltonian is restricted to
\begin{equation}
\begin{split}
\mathcal{H}_{\rm CF}(r, \theta, \varphi) 
=& r^2 q_{20} C_{20}(\theta, \varphi) + r^4 q_{40} C_{40}(\theta, \varphi) \\
&+ r^4 q_{43} (C_{43}(\theta, \varphi)-  C_{4-3}(\theta, \varphi)). 
\end{split}
\end{equation}
Considering the locations of ligands, we find 
\begin{equation}
\sum_{i=1}^6 Y^*_{km}(\theta_i, \varphi_i) = 6 Y_{km}(\theta', 0)
\end{equation}
for $(k,m)=(2,0), (4,0), (4, \pm 3)$, and we obtain 
\begin{equation}
\begin{split}
\mathcal{H}_{\rm CF}(r, \theta, \varphi) 
&= 6Ze^2 \bigg( \frac{r^2}{r_0^3}\frac{4\pi}{5} Y_{20}(\theta, \varphi) A_{20}(\theta') \\
&\quad+ \frac{r^4}{r_0^5}\frac{4\pi}{9}\sum_{m=0,\pm 3} Y_{4m}(\theta, \varphi) A_{4m}(\theta') \bigg), 
\end{split}
\end{equation}
where $A_{km}(\theta')=Y_{km}(\theta', 0)$. 
Using the $d$-orbitals basis set 
$\{ \psi_{xy}, \psi_{yz}, \psi_{zx}, \psi_{x^2-y^2} , \psi_{3z^2-r^2}\}$ in the local coordinate system 
the matrix representation of CF Hamiltonian is given by
\begin{widetext}
\begin{equation}\label{eq:trigonal_CF}
\mathcal{H}_{\rm CF} = 10Dq \mqty( (-3a_4-10a_2)/15 & -b/2 & 0 & 0 & 0 \\ -b/2 & (12a_4 +5a_2)/15 & 0 & 0 & 0 \\ 0 & 0 & (-3a_4-10a_2)/15 & b/2 & 0 \\ 0 & 0 & b/2 & (12a_4+5a_2)/15 & 0 \\ 0 & 0 & 0 & 0 & (-18a_4+10a_2)/15),
\end{equation}
\end{widetext}
where
\begin{equation}
\begin{split}
a_2 &= \frac{108\sqrt{5\pi}}{175} \kappa A_{20}(\theta') = \frac{27}{35}\kappa (3\cos^2\theta'-1),  \\
a_4 &= -\frac{4\sqrt{\pi}}{7} A_{40}(\theta') \\
&= -\frac{3}{2} \left( \frac{5}{2}\cos^4\theta' -\frac{15}{7}\cos^2\theta' +\frac{3}{14} \right), \\
b &= -8\sqrt{\frac{\pi}{35}} A_{43}(\theta') = 3\sin^3\theta'\cos\theta'. 
\label{eq:a2a4}
\end{split}
\end{equation}
Here we introduced the parameter $\kappa$, 
\begin{equation}
\kappa = r^2_0 \frac{\expval{r^2}}{\expval{r^4}} = r^2_0 \frac{\int_0^\infty r^2 R^2_{n2}(r) r^2 \dd r}{\int_0^\infty r^4 R^2_{n2}(r) r^2 \dd r}, 
\end{equation}
which is the parameter to include the screening effect through the radial wave function. 
The rough estimation and {\it ab initio} calculations give $\kappa=0.1\sim1$ \cite{Khomskii2015, Khomskii2016}. 
Since $\kappa>0$ by definition and $\theta'>\theta_0$ ($\theta'<\theta_0$) for the compression (stretching) of the octahedron, $a_2$ is positive/negative. 
Diagonalizing Eq.~(\ref{eq:trigonal_CF}), we obtain the eigenenergies
\begin{equation}
E_{a_{1g}} = \frac{-18a_4+10a_2}{15}, \quad E_{e_g^{\pi/\sigma}} = \frac{9a_4 -5a_2}{30} \mp \frac{\sqrt{a^2+b^2}}{2}, 
\end{equation}
where $-$/$+$ for $e^\pi_g$/$e^\sigma_g$, and $a=a_2 +a_4$. 
The corresponding eigenstates are expressed as
\begin{equation}\label{eq:trigonal_wf}
\begin{split}
\ket*{a_{1g}} &= \ket*{3z^2-r^2}, \\
\ket*{e_{g1}^\pi} &= -\cos\frac{\alpha}{2} \ket*{xy} +\sin\frac{\alpha}{2} \ket*{yz}, \\
\ket*{e_{g2}^\pi} &= \cos\frac{\alpha}{2} \ket*{x^2-y^2} +\sin\frac{\alpha}{2} \ket*{zx}, \\
\ket*{e_{g1}^\sigma} &= \sin\frac{\alpha}{2}\ket*{x^2-y^2} -\cos\frac{\alpha}{2} \ket*{zx}, \\
\ket*{e_{g2}^\sigma} &= -\sin\frac{\alpha}{2} \ket*{xy} -\cos\frac{\alpha}{2} \ket*{yz}, \\
\end{split}
\end{equation}
where $\cos\alpha=a/\sqrt{a^2 +b^2}$. 

%%%%%%%%%%%
Next, we introduce the atomic SOC that splits the energy level, following the treatment in Ref.~\cite{Khomskii2016}. 
Since the cubic CF energy splitting $\Delta_{\rm CF}=10Dq$ is large compared to the trigonal splitting $\Delta_1$ and atomic SOC constant $\zeta$, the upper $e_g^\sigma$ orbitals are discarded and the basis set is given as 
\{$\ket*{a_{1g}; \uparrow}$, $\ket*{e^\pi_{g+}; \uparrow}$, $\ket*{e^\pi_{g-}; \uparrow}$, $\ket*{a_{1g}; \downarrow}$, $\ket*{e^\pi_{g+}; \downarrow}$, $\ket*{e^\pi_{g-}; \downarrow}$\}, whose energy levels are shown in Fig.~\ref{fS9}(b) where
\begin{equation}\label{eq:trigonal_wf2}
\ket*{e^\pi_{g\pm}; \sigma} = \frac{1}{\sqrt{2}} (i \ket*{e^\pi_{g1}; \sigma} \mp \ket*{e^\pi_{g2}; \sigma}). 
\end{equation}
The matrix representation of Hamiltonian is given as
\begin{equation}
\mathcal{H} =  \mathcal{H}_{\rm CF} + \mathcal{H}_{\rm SOC} = \mqty( V & 0 \\ 0 & V) +\frac{\zeta}{2} \mqty( l_z & l_- \\  l_+ & -l_z ), 
\end{equation}
where
\begin{equation}
\begin{split}
V &=\mathrm{diag}(E_{a_{1g}}, E_{e_g^{\pi}}, E_{e_g^{\pi}}), \\
l_z &= \mathrm{diag}\left(0, \frac{1+3\cos\alpha}{2}, -\frac{1+3\cos\alpha}{2}  \right), \\
l_+ &= \mqty( 0 & -\sqrt{6}\sin\frac{\alpha}{2} & 0 \\ 0 & 0 & 0 \\ -\sqrt{6}\sin\frac{\alpha}{2} & 0 & 0 ). 
\end{split}
\end{equation}
Diagonalizing $\mathcal{H}$, we obtain three doublets with eigenenergies
\begin{equation}
\begin{split}
E_1 &= 2(\Delta_1 -\Delta)+E_{a_{1g}},\\
E_2 &= \Delta -\sqrt{\Delta^2 +\xi^2}+E_{a_{1g}}, \\
E_3 &= \Delta +\sqrt{\Delta^2 +\xi^2}+E_{a_{1g}}, 
\end{split}
\end{equation}
where
\begin{equation}
\Delta = \frac{\Delta_1}{2} +\frac{1+3\cos\alpha}{8} \zeta, \quad \xi = \sqrt{\frac{3}{2}} \sin\frac{\alpha}{2}\zeta. 
\end{equation}
The corresponding eigenstates are expressed as
\begin{equation}\label{eq:Kramers_doublets}
\begin{split}
\ket*{\phi^{(1)}_\uparrow} &= \ket*{e^\pi_{g-}; \uparrow}, \quad \ket*{\phi^{(1)}_\downarrow} = \ket*{e^\pi_{g+}; \downarrow},  \\
\ket*{\phi^{(2)}_\uparrow} &=\cos\delta\ket*{a_{1g}; \uparrow}+\sin\delta\ket*{e^\pi_{g-}; \downarrow}, \\
\ket*{\phi^{(2)}_\downarrow} &=\cos\delta\ket*{a_{1g}; \downarrow}+\sin\delta\ket*{e^\pi_{g+}; \uparrow}, \\
\ket*{\phi^{(3)}_\uparrow} &=\sin\delta\ket*{a_{1g}; \uparrow}-\cos\delta\ket*{e^\pi_{g-}; \downarrow}, \\
\ket*{\phi^{(3)}_\downarrow} &=\sin\delta\ket*{a_{1g}; \downarrow}-\cos\delta\ket*{e^\pi_{g+}; \uparrow}, 
\end{split}
\end{equation}
where the parameter $\delta$ is defined by
\begin{equation}
\tan 2\delta = \frac{\xi}{\Delta}, \quad \left( 0\leq \delta < \frac{\pi}{4},  \frac{\pi}{4} < \delta \leq \frac{\pi}{2} \right). 
\end{equation}
The range of values that $\delta$ takes is $0< \delta < \frac{\pi}{4}$ for $\Delta >0$ and $\frac{\pi}{4} < \delta < \frac{\pi}{2}$ for $\Delta<0$. 
In the limit of small SOC, $\zeta \ll \Delta_{\rm CF}$, $\delta\sim0$ for $\Delta >0$ and $ \delta \sim \frac{\pi}{2}$ for $\Delta <0$. 
In that case, $E_2$ ($E_3$) doublet behaves like $a_{1g}$ ($e_{g}^\pi$) orbitals for $\Delta >0$, and $E_3$ ($E_2$) doublet does for $\Delta <0$. 
Combining Eqs.~(\ref{eq:trigonal_wf}), (\ref{eq:trigonal_wf2}), (\ref{eq:Kramers_doublets}), and
\begin{equation}
\begin{split}
&\ket*{xy} = -\frac{i}{\sqrt{2}} (\ket*{l_z=2} - \ket*{l_z=-2}), \\
&\ket*{yz} = \frac{i}{\sqrt{2}} (\ket*{l_z=1} + \ket*{l_z=-1}), \\
&\ket*{zx} = -\frac{1}{\sqrt{2}} (\ket*{l_z=1} - \ket*{l_z=-1}), \\
&\ket*{x^2-y^2} = \frac{1}{\sqrt{2}} (\ket*{l_z=2} + \ket*{l_z=-2}), \\
&\ket*{3z^2-r^2} = \ket*{l_z=0}, 
\end{split}
\end{equation}
we can expand the above Kramers doublets by $\{\ket*{l_z,\sigma}\}$. 
In the following, we focus on one Kramers doublet depending on the number of $d$ electrons and consider it as a pseudospin $\Lambda=\Uparrow,\Downarrow$.

%%%%%%%%%%%
Finally, we represent the pseudospin in the {\it global} axis, performing local rotations of the coordinates to take the global $z_0$-axis as the quantization axis. 
As shown in Fig.~\ref{fS9}(c), the rotation axis $\hat{\vb*{n}_\mu}$ and the rotation angle $\varphi_\mu$ at each sublattice $\mu$ are given by $\hat{\vb*{n}}_{\mu} = \hat{\vb*{m}}_{\mu} \cross \vb*{e}_{z_0} /|\hat{\vb*{m}}_{\mu} \cross \vb*{e}_{z_0}|$ and $\varphi_\mu=\arccos(\hat{\vb*{m}}_{\mu} \cdot \vb*{e}_{z_0})$, respectively. 
The corresponding local rotations of the orbitals and spins are given by
\begin{equation}
\mathcal{R}_\mu (\varphi_\mu) = \exp \left[ +i\varphi_\mu \hat{\vb*{n}}_{\mu} \cdot \vb*{L} \right], 
\end{equation}
and
\begin{equation}
\mathcal{D}_\mu (\varphi_\mu) = \exp \left[ +i\frac{\varphi_\mu}{2} \hat{\vb*{n}}_{\mu} \cdot \vb*{\sigma} \right], 
\end{equation}
respectively. 
Here, $\vb*{L}=(L_x, L_y, L_z)$ is the orbital angular momentum operator. 
Then, we find the pseudospin at each sublattice $\mu$ given by
\begin{equation}\label{eq:local_trans_pseudospin}
\ket*{\Lambda_\mu} = \big(\mathcal{R}_\mu (\varphi_\mu) \otimes \mathcal{D}_\mu (\varphi_\mu)\big) \ket*{\Lambda}, 
\end{equation}
and obtain the coefficients $\{X^{\mu\Lambda_\mu}_{l_z,\sigma}\}$. 

%%%%%%%%%%%
We calculate the coefficients for $\rm CsW_2O_6$, where its pseudospin is the $E_2$ doublet. 
Its material parameters are $10Dq=2\, \rm eV$, $\theta_{\rm tri}=55.71^\circ$, and $\zeta=0.337\, \rm eV$, and we set $\kappa=1$. 
Expanding the pseudospin as
\begin{equation}
\ket*{\Lambda} = \sum_{l_z,\sigma} X^{\Lambda}_{l_z,\sigma} \ket*{l_z,\sigma}, 
\end{equation}
we find $X^{\pm'}_{\pm1, \mp}=-0.34$, $X^{\pm'}_{0, \pm}=0.82$, $X^{\pm'}_{\mp2, \mp}=\pm0.46$, and all others are zero, where $\Lambda=\Uparrow/\Downarrow$ for $+'/-'$ and $\sigma=\uparrow/\downarrow$ for $+/-$. 

%%%%%%%%%%%%%%% bibliography %%%%%%%%%%%%%%%
%%%%%%%%%%%%%%%%%%%%%%%%%%%%%%%%%%%%%%%%%%%%
\nocite{apsrev41Control}
\bibliographystyle{apsrev4-2}
\bibliography{flatband}
%%%%%%%%%%%%%%%%%%%%%%%%%%%%%%%%%%%%%

\end{document}